\documentclass[preprintnumbers,prd,showpacs,floatfix,superscriptaddress,nofootinbib,twocolumn]{revtex4-2}
\pdfoutput=1
\usepackage{graphicx,epsfig}
\usepackage{amsmath}
\usepackage {amssymb}
\usepackage {longtable}
\usepackage{multirow}
\usepackage{dcolumn}
\usepackage{bm}
\usepackage{amsfonts}
\usepackage{subfigure}
\usepackage{color}
\usepackage{relsize}

\newcommand{\m}{\mu}

\newcommand{\Si}{\Sigma}
\newcommand{\ch}{\mc{H}}

\newcommand{\mc}{\mathcal}
\newcommand{\ce}{\mc{E}}

\newcommand{\ca}{\mc{A}}

\begin{document}

\title{Junction conditions for LRS spacetimes in the $1+1+2$ covariant formalism}

\author{Jo\~{a}o Lu\'{i}s Rosa}
\email{joaoluis92@gmail.com}
\affiliation{Institute of Physics, University of Tartu, W. Ostwaldi 1, 50411 Tartu, Estonia}
\affiliation{University of Gda\'{n}sk, Jana Ba\.{z}y\'{n}skiego 8, 80-309 Gda\'{n}sk, Poland}

\author{Sante Carloni}
\email{sante.carloni@unige.it}
\affiliation{DIME Sez. Metodi e Modelli Matematici, Universit\`{a} di Genova,\\ Via All'Opera Pia 15, 16145 - Genoa, Italy.}
\affiliation{Institute of Theoretical Physics, Faculty of Mathematics and Physics,
Charles University, Prague, V Hole{\v s}ovi{\v c}k{\' a}ch 2, 180 00 Prague 8, Czech Republic}
\affiliation{INFN Sezione di Genova, Via Dodecaneso 33, 16146 Genova, Italy}

\date{\today}

\begin{abstract} 
We use the distribution formalism to derive the complete set of junction conditions for general Local Rotationally Symmetric (LRS) spacetimes in the $1+1+2$ covariant formalism. We start by developing a parametric framework encompassing timelike, spacelike, or null hypersurfaces. We then introduce the distribution formalism in the $1+1+2$ framework and obtain the necessary conditions to preserve the regularity of the $1+1+2$ equations at the separation hypersurface. Using these results, we can deduce some general prescriptions on the junction of LRS spacetimes and the properties of the shell in the non-smooth cases. As examples of the application of the junction conditions, we use this formalism to perform the matching necessary to obtain well-known solutions, e.g., the Martinez thin-shell, the Schwarzschild constant-density fluid star, and the Oppenheimer-Snyder collapse.
\end{abstract}

\maketitle

\section{Introduction}\label{sec:intro}

When dealing with astrophysically and cosmologically relevant spacetimes in the context of general relativity, one often faces the problem that global exact solutions cannot easily represent such spacetimes. This is particularly true in the case of the solution of relativistic compact stars, for which only a few approximated instances, e.g., the Vaidya spacetime, are known. 

A way to overcome this hurdle is to separate the complete spacetime into two or more sub-regions and then connect those regions. How this connection is performed was proposed for the first time by Darmois \cite{Darmois}, Lichnerowicz \cite{L1,L2}, and Israel \cite{Israel} (see also the works of Choquet--Bruhat \cite{Choquet} and Taub \cite{Taub}). The most commonly used of these formulations is undoubtedly the Israel junction conditions, which have the advantage of being formulated in terms of tensorial conditions. However, even if the solution to the problem is generally considered well established, much research is still being pursued on approaches that can overcome the limitations of the distributional approach \cite{Clarke1986,Huber:2019cze}  like, e.g., the variational definition of the junction conditions \cite{Mukohyama:2000ga,Barcelo:2000tga}. 

In general relativity (GR), junction conditions were proven useful in accounting for a wide range of astrophysical phenomena, e.g., the existence of fluid stars \cite{schwarzschildstar} and the Oppenheimer-Snyder collapse \cite{Oppenheimer:1939ue}. More recently, this formalism was also used in the context of exotic compact objects to derive stable and physically relevant solutions for black-hole mimickers \cite{rosafluid,rosafluid2,Tamm:2023wvn}. However, the set of junction conditions depends on the theory of gravity used as a framework. Thus, several works have also derived these conditions in extended theories of gravity, e.g. theories with additional scalar degrees of freedom \cite{Rosa:2023tph,senovilla1,Vignolo:2018eco,Reina:2015gxa,Deruelle:2007pt,Olmo:2020fri,rosafrt,rosafrt2,rosahmp,suffern, Barrabes:1997kk,Padilla:2012ze}, teleparallel theories of gravity \cite{delaCruz-Dombriz:2014zaa}, Einstein-Cartan theories of gravity \cite{Arkuszewski:1975fz}, and metric-affine gravity \cite{amacias}, with applications in wormhole physics \cite{rosaworm1,rosaworm2,rosaworm3}.

The aim of this paper is to provide a new formulation of junction conditions for Locally Rotationally Symmetric (LRS) spacetimes \cite{Ellis:1966ta,Stewart:1967tz}. This formulation is covariant, as the Israel-Darmois conditions, but it differs because of its primary focus: the existence of continuous congruences of worldlines crossing the boundary surface.  Such a different point of view is ascribed to the primary approach used to describe these conditions: the so-called {\it covariant approach}. This approach is based on the seminal work of Ehlers \cite{ehlers}, further developed by Ellis and other authors \cite{others1,others2,others3}. These methods rely on the threading of the spacetimes by means of specific vector fields associated with the motion of a given observer. 

There are two different realizations of the covariant approach. The first, called {\it 1+3 covariant approach}, is useful in the context of exact relativistic (and Newtonian) cosmological models \cite{Cargese,FLRW-pert}. A second one, which is also an extension of the 1+3 covariant approach, is the {\it 1+1+2 covariant approach} \cite{green,clarkbar,clarkson,betschart,Burston:2007wt}, which can be employed for the analysis of spacetimes of astrophysical interest \cite{nzioki,Carloni:2014rba,Bradley:1998tt,sante1,sante2,Luz:2019frs,Naidu:2021nwh}.

Since junction conditions are primarily employed in astrophysics, in this work we use the 1+1+2 approach. In addition, as the 1+3 formalism is entirely contained in the 1+1+2 one, using the latter results in no loss generality for our treatment.  We show that other than complementing the work in \cite{sante1,sante2,Luz:2019frs,Naidu:2021nwh} on the covariant formulation of the TOV equations, the relations we derive prove particularly useful in understanding some general properties of the junction conditions, and in the analysis of junction conditions in modifications of general relativity. Our work is not the first to present such an attempt. For example, junction conditions were given in the case of LRS-II spacetime using the 1+1+2 approach in \cite{Khambule:2020sgy}. In other words, e.g., the junction conditions are derived for specific cases/models (see, e.g., \cite{sante1,sante2, Bradley:1998tt}). However, in this work,  we take a slightly different approach in that we derive the general junction conditions directly from the distribution formalism and the 1+1+2 equations, extending our analysis to general LRS spacetimes and considering the case of non-comoving observers.  

This paper is organized as follows. Sec. \ref{sec:1+1+2Sec} introduces the 1+1+2 formalism and the respective equations.  In Sec. \ref{sec:1ParForm}, we develop a parametric formalism to describe several geometrical quantities, namely the metrics, hypersurfaces, and extrinsic curvature in the cases of timelike, spacelike, and null boundaries. In Sec. \ref{sec:formalism}, we introduce the formalism of distribution functions to derive a first set of junction conditions and the singular equations from which the remaining junction conditions arise. In Sec. \ref{sec:junctions}, we solve the singular equations and formulate the covariant junction conditions for LRS spacetimes. In Sec. \ref{sec:exs}, we provide applications of the previous framework to some interesting, well-known cases in GR. Finally, we trace our conclusions and future perspectives in Sec.~\ref{sec:concl}.

\section{The 1+1+2 formalism in LRS spacetimes}\label{sec:1+1+2Sec}
In this section, we introduce the 1+1+2 formalism for LRS spacetimes.  The 1+1+2 formalis is particularly advantageous in these spacetimes as all vector and tensor 1+1+2 quantities vanish. For brevity, in the following, we define and employ only those scalar 1+1+2 potentials, referring the reader to \cite{Cargese,clarkbar,clarkson} for additional details. 

\subsection{1+1+2 quantities and equations}

We start by defining two threading vector fields, $u^a$ and $e^a$. The first is time-like, i.e., $u^a u_a = -1$, and represents the four-velocity of an observer describing the spacetime. The second is space-like, i.e., $e_a e^a = 1$, which singles out a spatial direction for this observer. Using $u^a$ and $e^a$, we can define two projection tensors
\begin{eqnarray}
{h^a}_b &=& {g^a}_b + u^a u_b \hspace{0.2cm},\hspace{0.2cm} {h^a}_a = 3, \nonumber \\
{N_a}^b &=& {h_a}^b - e_a e^b = {g_a}^b + u_a u^b - e_a e^b \hspace{0.1cm},\hspace{0.1cm}{N^a}_a = 2,\hspace{0.6cm}\label{metricin112}
\end{eqnarray}
Where ${g^a}_b$ is the four-dimensional spacetime metric, ${h^a}_b$ represents the metric of the 3-spaces orthogonal to $u^a$, and ${N_a}^b$ represents the metric of the 2-spaces orthogonal to both $u^a$ and $e^a$\footnote{To be precise $h_{ab}$ and $N_{ab}$ are not the induced metrics, but rather. tensors defined in the whole manifold and that, on the surface, coincide with the induced metric. We use an index $H$ to distinguish the general tensors from the ones at the boundary. The same holds for all other quantities: index $H$ indicates their value at the boundary.\label{footnote1}}. 

Any tensorial object may be split according to the foliations given in Eq.\eqref{metricin112} into a set of quantities defined on these sub-spaces \cite{sante1}. For example, the covariant derivative $\nabla_a$ can be split into the covariant time derivative, the orthogonally projected covariant derivative, and the hat-derivative and $\delta$-derivative:
\begin{equation}\label{Cov_Der}
\begin{split}
{\dot{X}{}^{a..b}}_{c..d} &\equiv u^e {\nabla}_e {X^{a..b}}_{c..d}, \nonumber \\
D_e {X^{a..b}}_{c..d} &\equiv {h^a}_f...{h^b}_g{h^p}_c... {h^q}_d {h^r}_e {\nabla}_r {X^{f..g}}_{p..q}, \nonumber \\
{\hat{X}{}_{a..b}}^{c..d} &\equiv e^f D_f {X_{a..b}}^{c..d}, \nonumber \\
{\delta}_e {X_{a..b}}^{c..d} &\equiv {N_a}^f...{N_b}^g {N_i}^c...{N_j}^d {N_e}^p D_p {X_{f..g}}^{i..j}. 
\end{split}
\end{equation}

Using the operators given above, one can define the following relevant quantities for LRS spacetimes:
\begin{subequations} \label{variables112}
\begin{align} 
& \mathcal{A} = e_a \dot{u}^a, \hspace{0.2cm}\\
&\phi = {\delta}_a e^a, \\
& \theta=D_a u^a, \\
& \Sigma =  \frac{1}{3} D_a u_b\left(2 e^a e^b - N^{ab}\right)\\
& \Omega = \frac{1}{2} \varepsilon^{ab} \delta_{[a} u_{b]} \hspace{0.2cm},\\
&\xi = \frac{1}{2} \varepsilon^{ab} {\delta}_a e_b,  \\
& \mathcal{E} = {C^{ab}}_{cd} u_a u^d e_b e^c ,  \\
& \mathcal{H} = \frac{1}{2} {\varepsilon^a}_{de} {C^{deb}}_c u^c e_a e_b ,  
\end{align}
\end{subequations} 
where $\varepsilon_{abc} = \eta_{dabc} u^d$ and $\varepsilon_{ab} \equiv \varepsilon_{abc} e^c$ are the volume elements of the hypersurfaces perpendicular to $u^a$ and to both $u^a$ and $e^a$, respectively, $\eta_{abcd}$ is the Levi-Civita tensor, $C_{abcd}$ is the Weyl tensor, round parenthesis denote index symmetrization, square parenthesis denote index anti-symmetrization, i.e.,
\begin{equation}
X_{\left(ab\right)}=\frac{1}{2}\left(X_{ab}+X_{ba}\right),
\end{equation}
\begin{equation}
X_{\left[ab\right]}=\frac{1}{2}\left(X_{ab}-X_{ba}\right),
\end{equation}
Following the same projection procedure, the energy-momentum tensor $T_{ab}$ can be decomposed in the form
\begin{equation}\label{Tab}
\begin{split}
T_{ab} = &\mu u_a u_b + p_e e_a e_b \\
&+ p_N N_{ab}  + 2 Qu_{(a} e_{b)} .
\end{split}
\end{equation}
where 
\begin{equation}
\begin{split}
p_e &= p + \Pi,  \\
p_N &= p - \frac{1}{2} \Pi , 
\end{split}
\end{equation}
represent pressures along and orthogonal to $e_a$, respectively, and we have defined
\begin{equation}
\begin{split}
\mu &= T_{ab} u^a u^b,  \\
p &= \frac{1}{3} T_{ab} \left(e^a e^b + N^{ab}\right), \\
\Pi &= \frac{1}{3} T_{ab} \left(2 e^a e^b - N^{ab}\right), \\
Q &= -T_{ab} e^a u^b, 
\end{split}
\end{equation}
In the above equations, $\mu$ is the energy density, $p$ is the isotropic pressure, $\Pi$ represents the scalar component of the anisotropic pressure, and $Q$ represents the scalar part of the heat flux. Notice that in the previous equations, we are considering a geometrized unit system for which $8\pi G=c=1$.

 From the point of view of observers comoving with the fluid (if any\footnote{In the absence of sources, i.e., in vacuum spacetimes, $u^a$ remains undetermined, and, strictly speaking, the classification above does not apply. An exception is the (exterior of) Schwarzschild spacetime, which is empty but in which the observer's motion can be characterized in terms of the motion with respect to the central mass/event horizon. Indeed, in Schwarzschild's spacetime, a static observer is also a moving observer, and therefore we can classify it as LRS-II. For a general definition of observers and observables, we refer the interested reader to \cite{Bini:2014ndd}  and references therein.}) and that chooses $e^a$ to represent at each point the axes of symmetry of the metric, LRS spacetimes filled with a perfect fluid are naturally divided into three classes named LRS-I, LRS-II, and LRS-II. We call these observers ``comoving LRS observers''. 
 
In LRS-I spacetimes,  vorticity is absent from the spacelike congruence, i.e., $\xi=0$. We call this quantity "twist" to differentiate it from its counterpart $\Omega$ associated with the timelike congruence. These spacetimes do not expand and are shearless, i.e. $ \theta=0$, and $\Sigma=0$. The only non-zero quantities are thus $\phi$, $\ca$, and $\Omega$.  In LRS-II spacetimes, the vorticity terms and the magnetic part of the Weyl tensor vanish, i.e., $\Omega=0$, $\xi=0$, and $\mathcal{H}=0$. Every other scalar quantity may be non-zero. In the particular case of static and spherically symmetric LRS-II spacetimes, one also has $ \theta=0$, and $\Sigma=0$, and the dot derivative of every scalar vanishes. Finally, in LRS-III spacetimes, vorticity is absent from the timelike congruence, i.e., $\Omega=0$, and the only non-zero scalars are $ \ca$, $\theta$, $\Sigma$, $\xi$, $\mathcal E$, and $\mathcal H$. Further details of these metrics' structure and the solutions they encompass can be found in Appendix \ref{1+1+2 coord} and \cite{Ellis:1966ta,Stewart:1967tz}. Naturally, one can consider the case in which matter is not a perfect fluid, and the geometry is still one of the LRS subclasses. However, it is not necessarily true that the spacetime belongs globally to that LRS subclass. In the following, we assume that, for the cases we consider, the spacetime can always be characterized as LRS-I, LRS-II, or LRS-III regardless of the thermodynamics of the matter.

Any LRS spacetime can be fully described in terms of the following equations \cite{clarkson,betschart,Burston:2007wt}, which we refer to as the ``1+1+2 equations'':

{\it Evolution}
\begin{align}
& \dot\phi+\Bigl(\Si-\frac 23\, \theta\Bigr)\Bigl(\ca-\frac12\,\phi\Bigr)-2\,\xi\,\Omega =Q,\label{luphi}\\
& \dot\Si -\frac23\,\dot\theta -\frac12 \,\Bigl(\Si-\frac 23 \,\theta\Bigr)^2 +\ca\,\phi+\ce+2\,\Omega^2= \nonumber\\
&~~~~~~~~~~~~~~~~~~~= \frac13\,(\m+3\, p)+\frac12\,\Pi,\label{lusimthe}\\
&\dot\ce -\frac 13\dot\m+\frac 12\dot\Pi= \frac 32\,\Bigl( \Si-\frac23\theta\Bigr)\ce +\frac 14\Bigl( \Si-\frac23\theta\Bigr)\Pi\nonumber
\end{align}
\begin{align}
&~~~~~~~~~~~~~~+3\ch\xi +\frac 12 \phi Q-\frac 12(\m+p)\Bigl(\Si-\frac 23 \theta\Bigr),\label{luce}\\
&\dot\ch -\frac 32\,\Bigl(\Si-\frac23\theta\Bigr)\ch+3\,\ce\,\xi =Q\,\Omega +\frac 32\,\Pi\,\xi,\label{luch}\\
&\dot \xi -2\,\Bigl(\Si -\frac 16\,\theta\Bigr) \xi  =0 \label{luxi},\\
&\dot \Omega -\Bigl(\Si-\frac 23\,\theta \Bigr) \Omega-\ca\,\xi=0 \label{luOmega} .
\end{align}
{\it Propagation} 
\begin{align}
&\hat\phi+\frac 12\,\phi^2-\Bigl(\Si-\frac 23\,\theta\Bigr)\Bigl(\Si+\frac 13\,\theta\Bigr)+\ce-2\,\xi^2\nonumber\\
&~~~~~~~~~~~~~~~~~~~=-\frac 23\,\mu -\frac 12\,\Pi, \label{rb1}\\
&\hat\Si -\frac23\,\hat\theta+\frac 32\, \phi\,\Si+2\,\xi\,\Omega=-Q,\label{lensimthe}\\
&\hat \ce-\frac13\, \hat \m+\frac 12\,\hat \Pi =3\,\ch\,\Omega+\frac 12\,\Bigl( \Si-\frac23\,\theta\Bigr)\,Q\nonumber\\
&~~~~~~~~~~~~~~~~~~~-\frac 32\,\phi\,\ce -\frac 34\,\phi \,\Pi,\label{lence}
\end{align} 
\begin{align}
&\hat \ch +\frac 32\,\phi\,\ch+3\,\ce\,\Omega=-\Bigl(\m+p-\frac 12\,\Pi\Bigr)\,\Omega-Q\,\xi,\\
&\hat \xi +\phi\, \xi -\Bigl(\Si+\frac 13\,\theta\Bigr)\Omega =0\label{lenxi}, \\
&\hat \Omega -(\ca-\phi) \,\Omega =0 .\label{lenomega}
\end{align} 
{\it Mixed}
\begin{align}
&\hat \ca+(\ca+\phi)\ca-\dot\theta-\frac 13\, \theta^2-\frac 32\, \Si^2+2\,\Omega^2\nonumber\\
&~~~~~~~~~~~~~~~~~~~~~= \frac 12 \,(\m+3\,p),\label{lusidas}\\
&\dot\m +\hat Q+(2\,\ca+\phi)Q+ \theta\,(\m + p)+\frac 32\, \Si\, \Pi =0,\label{tabcons}\\
&\dot Q+\hat p+\hat \Pi +\ca\,(\m+p+\Pi ) +\nonumber\\
&~~~~~~~~~~~~~~~~~+\frac 32\,\phi\Pi+\Bigl(\Si+\frac 43 \, \theta\Bigr)Q  =0.\label{rblast}
\end{align}
{\it Constraint}
\begin{equation}
3\,\xi\,\Si -(2\,\ca-\phi) \Omega -\ch=0. \label{heqs}
\end{equation}  

Finally, the Gauss curvature $K$ can be defined in terms of the 1+1+2 quantities as
\begin{align}
&K:=\frac 13\,\m-\ce-\frac 12 \,\Pi+\frac 14 \,\phi^2\nonumber\\
&~~~~~~~~~~~~~~~-\frac 14\,\Bigl(\Sigma-\frac 23\,\theta\Bigr)^2 +\xi^2-\Omega^2,\label{gauscurv}
\end{align}
 and it satisfies the evolution and propagation equations
\begin{equation} \label{Keq1}
\dot{K}=-\left(\frac{2}{3}\theta-\Sigma\right) K,
\end{equation}
\begin{equation} \label{Keq2}
\hat{K}=-\phi K,
\end{equation}
respectively. These two equations are not independent of the 1+1+2 system but are potentially valuable as auxiliary equations in several contexts, e.g., the covariant TOV equations \cite{sante1,sante2,Luz:2019frs,Naidu:2021nwh}. 

Finally, for any scalar quantity $X$ in a LRS spacetime, the following condition applies,
\begin{equation}
\varepsilon^{ab}\nabla_{a}\nabla_{b}X=0
\end{equation}  
which in turn implies
\begin{equation}
\Omega \dot{X}=\xi \hat{X}.
\end{equation}
Upon replacing all the 1+1+2 scalar potentials into the equation above, one verifies that for only three of these scalars, one generates non-trivial results, which imply the following three additional constraints:
\begin{equation}\label{Other_constr}
\begin{split}
&3\phi \xi +\Omega (3\Sigma-2\theta)=0,\\
& Q (\xi^2+\Omega^2)-(p+\Pi+\mu)\xi \Omega=0,\\
&\frac{3}{2} \xi  \Sigma  \phi+\xi  Q-2
   \Omega ^3+ \Omega  \left[\frac{1}{2}\Pi +\frac{1}{3}\mu +p-\mathcal{E}+\xi ^2
 \right.\\ 
&~~~~~~~~~~~~~~~~~~\left. -\mathcal{A} \phi+\frac{1}{2}\left(\frac{2}{3} \theta -3\Sigma \right)^2\right]=0.
\end{split}
\end{equation}
In the following, we deduce the junction conditions in an entirely covariant form in terms of the quantities defined in Eq. \eqref{variables112}. 

\section{A parametric formalism}\label{sec:1ParForm}

In order to obtain the general junction conditions, we now proceed to construct a parametric formalism for junction conditions. In particular, we provide a unified definition of the metric, the induced metric, and the extrinsic curvature of the separation hypersurface that characterizes the junction of two spacetimes for the timelike, spacelike, and null cases. 

\subsection{Normal vectors and induced metric} \label{NormMet}

Let us start by stating the geometrical quantities we want to obtain for each relevant case. 

Since the covariant approaches are based on foliations, it is convenient to construct, whenever possible, the junction conditions so that the separation hypersurface coincides with the foliation induced by the choice of $u^a$ and $e^a$. In this way, when one deals with timelike or spacelike hypersurfaces, it is sufficient to make use of the projected 2-metric $N_{ab}$, and the projection vectors $u_a$ and $e_a$ defined previously in Eq.\eqref{metricin112}. 
In this way, for the case of a timelike normal, the induced metric reads\footnote{Here, the same discussion of Foontote \ref{footnote1} holds. The tensor $q_{ab}$ is not strictly the induced metric but coincides with it on the separation surface. The same happens for the extrinsic curvature. However, as well known \cite{Poisson}, the projection operators associated with the pullback are continuous across the boundary, and, thus, the junction conditions can also be given in terms of $q_{ab}$.} 
\begin{equation}
    q _{ab}=h_{ab}= N _{ab}-u_{a} u_{b} .
\end{equation}
whereas, in the case of  spacelike normal
\begin{equation}
q _{ab}= N_{ab}+e_{a} e_{b}.
\end{equation}
If one is interested in null hypersurfaces, instead, the 4-metric $g_{ab}$ can be decomposed into the 2-metric $N_{ab}$ and the null vectors $l_a$ and $\bar l_a$ as
\begin{equation}\label{defgabnull}
g_{ab}=N_{ab}-l_a\bar l_b-\bar l_al_b,
\end{equation}
where both the unit vectors $l_a$ and $\bar l_a$ are null and satisfy the following properties
\begin{equation}\label{defvecll}
l_a=\frac{\sqrt{2}}{2}\left(u_a+e_a\right), \quad \bar l_a=\frac{\sqrt{2}}{2}\left(u_a-e_a\right), 
\end{equation}
\begin{equation}\label{definnerll}
l_al^a=0,\quad \bar l_a\bar l^a=0,\quad l_a\bar l^a=-1.
\end{equation}
For null hypersurfaces and the metric decomposition described above, the induced metric $q_{ab}$ on the hypersurface is the same as the 2-metric $N_{ab}$, i.e.
\begin{equation}\label{defhabnull}
q_{ab}=g_{ab}+l_a\bar l_b+\bar l_al_b=N_{ab},
\end{equation}
and both the vectors $l_a$ and $\bar l_a$ are orthogonal to the hypersurface.

With these relations in mind, and as it is explained in detail in Appendix \ref{Normal}, the normal vector $n_a$ to the hypersurface can be written in general  as
\begin{equation}\label{defnormgen}
n_a=\tau u_a+\varsigma e_a,
\end{equation}
in such a way that
\begin{equation}\label{sigtauchoice}
\left\{\begin{array}{ccccl}
&\tau=1& \varsigma=0 &\Rightarrow &\mbox{spacelike hypersurfaces,}\\
&\tau=0& \varsigma=1 &\Rightarrow &\mbox{timelike hypersurfaces,}\\
&\tau=\frac{\sqrt{2}}{2}&\varsigma=\pm\frac{\sqrt{2}}{2}&\Rightarrow &\mbox{null hypersurfaces.}\\
\end{array}\right.
\end{equation}
It would be tempting to assume, at this point, to consider  $\tau$ and $\varsigma$ as generic functions. However, this would lead to boundary hypersurfaces with a mixed character, which we do not consider in this work.

The choice of Eq. \eqref{defnormgen} presents a great generality insofar as the formalism is covariant and, therefore, there is no real constraint on the choice of the vectors $u_a$ and $e_a$. This is also true for the case of null boundary hypersurfaces. However, additional care is required to introduce vorticity and/or twist. For an LRS-I spacetime, for example, since the  $u_a$ associated with the comoving observers possesses a solenoidal part, it cannot be aligned to the normal vector to a boundary surface. A similar problem arises with $e_a$ in LRS-III with spacelike surfaces. Therefore, it is impossible in these cases to choose a normal aligned to $u_a$ or $e_a$.  Due to this issue, tn the following, we restrict our analysis to the cases of spacelike surfaces in LRS-I and LRS-II,  timelike surfaces in LRS-II and LRS-III,  and null surfaces in LRS-II only. 

\section{Distribution functions in the 1+1+2 formalism}\label{sec:formalism}

Another cornerstone of our construction is the distribution formalism (see, e.g., Ref.\cite{Poisson,Steinbauer:2006qi,Reina:2015gxa}), which we now summarize briefly to use as a framework on which to derive the junction conditions. 

\subsection{Type I junction conditions}
Consider a spacetime $\mathcal V$ divided into two parts, an exterior region $\mathcal V^+$ described by a metric $g_{ab}^+$, and an interior region $\mathcal V^-$ described by a metric $g_{ab}^-$. The two regions $\mathcal V^\pm$ are separated by a 3-dimensional hypersurface $H$ described by an induced metric $q^H_{ab}$. The projection vectors from the 4-dimensional spacetime $\mathcal V$ to the hypersurface $H$  can be obtained directly from the induced metric $q^H_{ab}$ as $(q^H)^{a}{}_{b}$. The displacement from $H$ is measured along curves tangent to the vector field $n_a$ normal to $H$ and is locally parametrized by an affine parameter $\ell$ by the relation
\begin{equation}
n_a=\varepsilon\partial_a \ell
\end{equation}
Without loss of generality, we choose $\ell$ to be zero at $H$, negative in the region $\mathcal V^-$ and positive in the region $\mathcal V^+$.

Any tensorial quantity in this setting can be generalized to the following distribution\footnote{For a more formal definition of the passage between tensorial functions and distributions, we refer the reader to \cite{senovilla1}.}
\begin{equation}\label{quantdist2}
\begin{split}
    X^{a...}{}_{b...}=&(X^{a...}{}_{b...})^+\Theta\left(\ell\right)+(X^{a...}{}_{b...})^-\Theta\left(-\ell\right)\\&+\bar X^{a...}{}_{b...}\delta\left(\ell\right),
\end{split}
\end{equation}
where $(X^{a...}{}_{b...})^\pm$ represents the quantity $X^{a...}{}_{b...}$ in the region $\mathcal V^\pm$,  $\Theta\left(\ell\right)$ and $\delta\left(\ell\right)$ are the Heaviside and Dirac delta distributions respectively, and $\bar X^{a...}{}_{b...}$  is the singular hypersurface component.  We also denote, as customary, the discontinuity of a quantity $X$ across $H$ and the surface value of $X$ as, respectively,
\begin{equation}
    \left[X^{a...}{}_{b...}\right]_{\pm}=(X^{a...}{}_{b...})^+_H-(X^{a...}{}_{b...})^-_H.
\end{equation} 
\begin{equation}
\langle X^{a...}{}_{b...}\rangle=\frac{1}{2}\left[\left(X^{a...}{}_{b...}\right)^+_H+\left(X^{a...}{}_{b...}\right)^-_H\right].
\end{equation} 
The quantity $ \left[X^{a...}{}_{b...}\right]_{\pm}$ is usually referred  to as the ``jump of $X^{a...}{}_{b...}$''. Taking the partial derivative of Eq. \eqref{quantdist2}, one obtains, using the fact that $\Theta'\left(\ell\right)=\delta\left(\ell\right)$,
\begin{equation}\label{partdist2}
\begin{split}
    \partial_r (X^{a...}{}_{b...})=&(\partial_r X^{a...}{}_{b...})^+\Theta\left(\ell\right)+\\
   &(\partial_r X^{a...}{}_{b...})^-\Theta\left(-\ell\right)+\\
   &\epsilon n_r\left[X^{a...}{}_{b...}\right]_{\pm}\delta\left(\ell\right)+\partial_r\left[\bar X^{a...}{}_{b...}\delta\left(\ell\right)\right] .
\end{split}
\end{equation}
Junction conditions can only be defined if the boundaries between the two spacetimes are correctly identified. We then identify the points of the hypersurface $H$ requiring that the normal vector field is continuous across $H$
\begin{equation}\label{n-n}
[n_a ]_\pm=0.
\end{equation}
This condition is also necessary for constructing a consistent reference frame across $H$, a requirement for the construction of the 1+1+2 junction conditions as well as for the Israel-Darmois conditions. At the same time, we have to ensure that the tangent spaces of the two spacetimes are correctly identified.  It has been shown, in general (see \cite{Clarke1986}), that  this is possible if Eq. \eqref{n-n} holds together with
\begin{equation}\label{JC_1}
    \left[q_{ab}\right]_\pm=0.
\end{equation}
We dub the above relation with the condition in Eq. \eqref{n-n}, type I junction conditions. These are mandatory conditions in any metric theory of gravity, and they arise directly from applying the distribution formalism. The remaining junction conditions, which we call type II conditions, are obtained by applying the same procedure to the 1+1+2 evolution, propagation, mixed, and constraint equations.

Let us start considering the consequences of these requirements on the 1+1+2 quantities. As $q_{ab}$ is obtained by the combination of the orthogonal quantities $N_{ab}$ and the tensor product of a vector orthogonal to $n_a$, the type I junction conditions  imply that
\begin{equation}\label{1+1+2ueN}
\begin{split}
u^{a}&=(u^{a})^+\Theta\left(\ell\right)+(u^{a})^-\Theta\left(-\ell\right),\\
e^{a}&=(e^{a})^+\Theta\left(\ell\right)+(e^{a})^-\Theta\left(-\ell\right),\\
N_{ab}&=(N^{ab})^+\Theta\left(\ell\right)+(N^{ab})^-\Theta\left(-\ell\right).
\end{split}
\end{equation}
In addition, Eq. \eqref{JC_1} applied to the definitions of the variables associated with the electric and magnetic part of the Weyl tensor shows that they are all well-defined in terms of distributions. This can be seen explicitly noting that Eq. \eqref{JC_1} implies that the Christoffel symbols $ \Gamma^c_{ab}$ are well defined in a distributional sense, and that $\mathcal{E}$ and $\mathcal{H}$ are functions of $ \Gamma^c_{ab}$, $u^{a}$, $e^{a}$, and $N_{ab}$. This result shows the actual reason why the Israel junction conditions work, even if there is no direct mention of the conformal structure of the spacetime. Equation \eqref{JC_1} is enough to guarantee that no pathologies arise in the Weyl tensor.
 
The rest of the 1+1+2 potentials are expressed in terms of the derivatives of the quantities defined in Eq. \eqref{1+1+2ueN}. Let us then look at how these can be well defined in terms of distributions. Since Eq. \eqref{JC_1} also implies that $\Gamma$ does not have a singular part, we can apply the principle of general covariance to Eq. \eqref{partdist2}. Furthermore, using the properties of the derivatives of the Dirac-$\delta$ distribution (see Appendix \ref{sec:APPdeltaderiv} for details), one obtains:
\begin{equation}\label{precovdist2}
\begin{split}
    \nabla_r (X^{a...}{}_{b...})=&(\nabla_r X^{a...}{}_{b...})^+\Theta\left(\ell\right)+(\nabla_r X^{a...}{}_{b...})^-\Theta\left(-\ell\right)\\
   +&\delta\left(\ell\right)\left(\epsilon n_r\left[X^{a...}{}_{b...}\right]_\pm+\nabla_r\bar{X}^{a...}{}_{b...}\right.\\
   -&\left.\epsilon n_r\left<\mathcal{K}\right>\bar{X}^{a...}{}_{b...}\right)+\bar X{\Delta^{a...}{}_{rb...}}\left(\ell\right),
\end{split}
\end{equation}
where $\mathcal{K}$ is the trace of the extrinsic curvature $\mathcal{K}_{ab}$ (see Appendix \ref{ExtCurv112} for details on its form in the 1+1+2 approach)  and ${\Delta^{a...}{}_{rb...}}\left(\ell\right)$ is a double gravitational layer distribution term associated with $\bar{X}^{a...}{}_{b...}$ \cite{senovilla1,Reina:2015gxa}. 
Then, Eq. \eqref{n-n} implies, for any quantity $X^{a...}{}_{b...}$, 
\begin{equation}\label{dotdist}
\begin{split}
\dot{X}^{a...}{}_{b...}=& (\dot X^{a...}{}_{b...})^+\Theta\left(\ell\right)+(\dot X^{a...}{}_{b...})^-\Theta\left(-\ell\right)\\
+&\delta\left(\ell\right) \left(\tau \left[X^{a...}{}_{b...}\right]_{\pm}+\dot {\bar X}^{a...}{}_{b...}\right. \\
-&\left.\tau\left<\mathcal{K}\right>\bar{X}^{a...}{}_{b...}\right)+u^r\bar X{\Delta^{a...}{}_{rb...}}\left(\ell\right),
\end{split}
\end{equation}
\begin{equation}\label{hatdist}
\begin{split}
\hat{X}^{a...}{}_{b...}=&(\hat X^{a...}{}_{b...})^+\Theta\left(\ell\right)+( \hat X^{a...}{}_{b...})^-\Theta\left(-\ell\right)\\
+&\delta\left(\ell\right) \left(\varsigma \left[X^{a...}{}_{b...}\right]_{\pm}+\hat {\bar X}^{a...}{}_{b...}\right.\\
-&\left.\varsigma\left<\mathcal{K}\right>\bar{X}^{a...}{}_{b...}\right)+e^r\bar X{\Delta^{a...}{}_{rb...}}\left(\ell\right),
\end{split}
\end{equation}
\begin{equation}\label{deltadist}
\begin{split}
\delta_r{X}^{a...}{}_{b...}=&(\delta_r X^{a...}{}_{b...})^+\Theta\left(\ell\right)+( \delta_rX^{a...}{}_{b...})^-\Theta\left(-\ell\right) \\
+&\delta_r\bar X^{a...}{}_{b...}+N_r{}^q\bar X{\Delta^{a...}{}_{qb...}}\left(\ell\right).
\end{split}
\end{equation}
Using the above results, one can prove that many of the 1+1+2 scalars do not have a singular part. For example, in the case of $\phi$, we have
\begin{equation}
\begin{split}
\phi =& \delta_a e ^a= N^{ab}\nabla_a e_b=\\
=& (N^{ab}\nabla_a e_b)^+\,\Theta\left(\ell\right)+(N^{ab}\nabla_a e_b)^-\,\Theta\left(-\ell\right)\\
&+ \epsilon N_{ab} n^a[e^b]_{\pm}\delta\left(\ell\right)
\end{split}
\end{equation}
Since $N_{ab}e^b=0$ in $H$, one concludes that the term proportional to $\delta\left(\ell\right)$ in $\phi$ vanishes independently of the form of $n^a$, and $\phi$ does not have a singular part. In the case of $\theta$ one has
\begin{equation}
\begin{split}
\theta =& D_a u ^a= h^{ab}\nabla_a u_b=\\
=& (h^{ab}\nabla_a u_b)^+\Theta\left(\ell\right)+(h^{ab}\nabla_a u_b)^-\Theta\left(-\ell\right)\\
&+ \epsilon h_{ab} n^a[u^b]_{\pm}\delta\left(\ell\right)
\end{split}
\end{equation}
Since we have $h_{ab}u^b=0$ in $H$ by construction, the term proportional to $\delta\left(\ell\right)$ in $\theta$ vanishes, and this quantity is regular. Similar arguments can be carried out for the variables $\mathcal A$, $\Sigma$, $\Omega$, and $\xi$. We then conclude that of all the 1+1+2 quantities, only the matter variables $\mu$, $p$, $Q$ and $\Pi$  and $\mathcal E$, $\mathcal H$ (and therefore $K$), can have a singular part. We then denominate $\bar \mu$, $\bar p$, $\bar \Pi$, $\bar Q$, $\bar{\mathcal E}$, $\bar{\mathcal H}$, and $\bar K$ the singular parts of these quantities.

The singular terms for matter can be associated with the presence of a thin shell of matter at the separation hypersurface, described by a stress-energy tensor $S_{ab}$. Indeed, the distributional form of the stress-energy tensor $T_{ab}$ of the matter sector can be written as\footnote{We note that, in general, the stress-energy tensor $T_{ab}$ may feature additional non-tangential singular terms associated with the double gravitational layer and external fluxes and tensions \cite{senovilla1,Reina:2015gxa,Rosa:2023tph}. Nevertheless,  since we will assume matter to be an incoherent fluid and work in General Relativity, we have excluded these terms from our definition of the stress-energy tensor.}
\begin{equation}\label{tabwithsab}
T_{ab}=T_{ab}^+\Theta\left(\ell\right)+T_{ab}^-\Theta\left(-\ell\right)+\delta\left(\ell\right)S_{ab}.
\end{equation}
where $S_{ab}$ represents the energy-momentum tensor of the shell. The matter quantities of the thin shell, i.e., the singular parts of the matter quantities, are thus connected to $S_{ab}$ by the following relations
\begin{eqnarray}\label{Sab}
\nonumber S_{ab} = &&\bar \mu u_a u_b + (\bar  p + \bar \Pi) e_a e_b \\
 &&+ 2 \bar Q u_{(a} e_{b)}+ \left(\bar p - \frac{1}{2} \bar \Pi \right) N_{ab},
\end{eqnarray}
or
\begin{eqnarray}
\bar\mu &=& S_{ab} u^a u^b, \nonumber \\
\bar p &=& \frac{1}{3} S_{ab} \left(e^a e^b + N^{ab}\right), \nonumber \\
\bar \Pi &=& \frac{1}{3} S_{ab} \left(2 e^a e^b - N^{ab}\right), \nonumber \\
\bar Q &=& \frac{1}{2} S_{ab} e^a u^b.
\end{eqnarray}
and, naturally, $S_{ab}n^b=0$.

Writing the stress-energy tensor in the form of Eq. \eqref{tabwithsab} and taking a covariant derivative using Eq. \eqref{precovdist2}, one obtains
\begin{equation}
\begin{split}
    \nabla_a T^{ab}=&\left(\nabla_a T^{ab}\right)^+\Theta\left(\ell\right)+\left(\nabla_a T^{ab}\right)^-\Theta\left(\ell\right)+\\
    +&\left(\epsilon n_a\left[T^{ab}\right]+D_a S^{ab}\right)\delta\left(\ell\right)\\
\end{split}
\end{equation}
The singular part of this equation is thus 
\begin{equation}\label{Cons_Shell}
D_a S^{ab}+\epsilon n_a\left[T^{ab}\right]=0
\end{equation}
 which represents the (non-)conservation laws of matter on the boundary. Notice that the above equation shows that the tensor $S_{ab}$ is not necessarily conserved, and therefore, in general, the boundary can itself be dynamic.

\subsection{Singular equations and constraints: type II junction conditions}\label{SinEqJC}

In analogy with the classical Israel derivation, we now deduce conditions for which the 1+1+2 equations are regular in the presence of a boundary surface. This is accomplished by determining and setting to zero the singular parts proportional to $\delta\left(\ell\right)$ and $\Delta_a\left(\ell\right)$, separately. We then obtain the following relations, from which the ``type II junction conditions'' can be extracted. For the terms proportional to $\delta(\ell)$, we obtain:

\textit{Evolution}
\begin{equation}\label{aQ}
 \tau\left[\phi\right]_{\pm}=\bar Q,
\end{equation}
\begin{equation}
\tau\left(\left[\Sigma\right]_{\pm}-\frac{2}{3}\left[\theta\right]_{\pm}\right)+\bar{\mathcal E}=\frac{1}{3}\left(\bar \mu+3\bar p\right)+\frac{1}{2}\bar\Pi,
\end{equation}
\begin{multline}
 \tau\left(\left[\mathcal E\right]_{\pm}-\frac{1}{3}\left[\mu\right]_{\pm}+\frac{1}{2}\left[\Pi\right]_{\pm}\right)-\frac{3}{2}\left(\langle\Sigma\rangle-\frac{2}{3}\langle\theta\rangle\right)\bar{\mathcal E}=\\
=-3\bar{\mathcal H}\langle\xi\rangle+\left(\langle\Sigma\rangle-\frac{2}{3}\langle\theta\rangle\right)\left[\frac{1}{4}\bar\Pi-\frac{1}{2}\left(\bar \mu + \bar p\right)\right]+\frac{1}{2}\langle\phi\rangle\bar Q,
\end{multline}
\begin{equation}
 \tau\left[\mathcal H\right]_{\pm}-\left(\langle\Sigma\rangle-\frac{2}{3}\langle\theta\rangle\right)\bar{\mathcal H}+3\bar{\mathcal E}\langle\xi\rangle=\langle\Omega\rangle\bar Q+\frac{3}{2}\langle\xi\rangle\bar\Pi,
\end{equation}
\begin{equation}\label{axi}
 \tau\left[\xi\right]_{\pm}=0,
\end{equation}
\begin{equation}\label{aomega}
 \tau\left[\Omega\right]_{\pm}=0.
\end{equation}

\textit{Propagation}
\begin{equation}\label{propphi}
\varsigma\left[\phi\right]_{\pm}+\bar{\mathcal E}=-\frac{2}{3}\bar\mu-\frac{1}{2}\bar\Pi,
\end{equation}
\begin{equation}\label{bQ}
\varsigma\left(\left[\Sigma\right]_{\pm}-\frac{2}{3}\left[\theta\right]_{\pm}\right)=-\bar Q,
\end{equation}
\begin{multline}
\varsigma\left(\left[{\mathcal E}\right]_{\pm}-\frac{1}{3}\left[\mu\right]_{\pm}+\frac{1}{2}\left[\Pi\right]_{\pm}\right)+\frac{3}{2}\langle\phi\rangle\bar{\mathcal E}-3\bar{\mathcal H}\langle\Omega\rangle=\\
=\frac{1}{2}\left(\langle\Sigma\rangle-\frac{2}{3}\langle\theta\rangle\right)\bar Q-\frac{3}{4}\langle\phi\rangle\bar\Pi,
\end{multline}
\begin{equation}
\varsigma\left[\mathcal H\right]_{\pm}+\frac{3}{2}\langle\phi\rangle\bar{\mathcal H}+3\bar{\mathcal E}\langle\Omega\rangle=-\langle\Omega\rangle\left(\bar \mu+\bar p-\frac{1}{2}\bar\Pi\right)-\langle\xi\rangle\bar Q,
\end{equation}
\begin{equation}\label{bxi}
\varsigma\left[\xi\right]_{\pm}=0,
\end{equation}
\begin{equation}\label{bomega}
\varsigma\left[\Omega\right]_{\pm}=0.
\end{equation}

\textit{Mixed}
\begin{equation}\label{propa}
\varsigma\left[\mathcal A\right]_{\pm}- \tau\left[\theta\right]_{\pm}=\frac{1}{2}\left(\bar \mu+3\bar p \right),
\end{equation}
\begin{multline}\label{JuncFin2}
 \tau\left(\left[\mu\right]_{\pm}+\dot{\bar\mu}\right)+\varsigma\left(\left[Q\right]_{\pm}+\hat{\bar Q}\right)=\\
=-\langle\theta\rangle\left(\bar \mu+\bar p\right)-\frac{3}{2}\langle\Sigma\rangle\bar \Pi-\left(2\langle\mathcal A\rangle+\langle\phi\rangle\right)\bar Q,
\end{multline}
\begin{multline}\label{JuncFin}
\tau\left(\left[Q\right]_{\pm}+\dot{\bar Q}\right)+\varsigma\left(\left[p\right]_{\pm}+\left[\Pi\right]_{\pm}+\hat{\bar p}+\hat{\bar \Pi}\right)=\\
=-\mathcal A\left(\bar\mu+\bar p\right)-\left(\langle\mathcal A\rangle+\frac{3}{2}\langle\phi\rangle\right)\bar\Pi-\left(\langle\Sigma\rangle+\frac{4}{3}\langle\theta\rangle\right)\bar Q.
\end{multline}
\textit{Constraints}
\begin{align}
\bar{\mathcal H}=0 \label{solweylH}.
\end{align}

On the other hand, for the terms proportional to $\Delta_a\left(\ell\right)$, we obtain:\\
\textit{Evolution:}
\begin{equation}\label{eqD:1}
    \tau\left(\bar{\mathcal E}-\frac{1}{3}\bar\mu+\frac{1}{2}\bar\Pi\right)=0,
\end{equation}
\begin{equation}
    \tau\bar{\mathcal H}=0.
\end{equation}
\textit{Propagation:}
\begin{equation}
    \varsigma\left(\bar{\mathcal E}-\frac{1}{3}\bar\mu+\frac{1}{2}\bar\Pi\right)=0,
\end{equation}
\begin{equation}\label{eqD:2}
    \varsigma\bar{\mathcal H}=0.
\end{equation}
\textit{Mixed:}
\begin{equation}\label{eqD:3}
    \tau\bar\mu+\varsigma\bar Q=0,
\end{equation}
\begin{equation}\label{eqD:4}
    \tau\bar Q+\varsigma\left(\bar p + \bar \Pi\right)=0.
\end{equation}

We also provide, for completeness, the evolution and propagation equations for the Gauss curvature $K$, given in Eqs. \eqref{Keq1} and \eqref{Keq2}. The singular parts of these equations proportional to $\delta\left(\ell\right)$ become, in this formalism:
\begin{equation}\label{prejunK1}
 \tau\left[K\right]_{\pm}+\dot{\bar K}=\left(\langle\Sigma\rangle-\frac{2}{3}\langle\theta\rangle\right)\bar K+\tau\left<\mathcal{K}\right>\bar K,
\end{equation}
\begin{equation}\label{prejunK2}
\varsigma\left[K\right]_{\pm}+\hat{\bar K}=-\langle\phi\rangle \bar K+\varsigma\left<\mathcal{K}\right>\bar K,
\end{equation}
whereas the singular parts proportional to $\Delta_a\left(\ell\right)$ immediately set
\begin{equation}\label{eq:barKzero}
    \bar K=0.
\end{equation}
As a consequence of Eq.\eqref{eq:barKzero}, one finds that also $\dot{\bar K}=\hat{\bar K}=0$ which, upon replacing back into Eqs. \eqref{prejunK1} and \eqref{prejunK2}, immediately imply that the Gauss curvature must be continuous across $H$, i.e.,
\begin{equation}
    \left[K\right]_\pm=0,
\end{equation}
independently of the character of the junction hypersurface. Furthermore, taking the singular part of Eq. \eqref{gauscurv} and using the fact that $\bar K=0$, one obtains the following constraint on the quantities $\bar\mu$, $\bar\Pi$ and $\bar{\mathcal{E}}$:
\begin{equation}\label{solweylE}
    \bar{\mathcal{E}}=\frac{1}{3}\bar\mu-\frac{1}{2}\bar\Pi.
\end{equation}

The constraints in Eqs. \eqref{solweylH} and \eqref{solweylE} guarantee that the evolution and propagation equations proportional to $\Delta_a\left(\ell\right)$, i.e., Eqs. \eqref{eqD:1} to \eqref{eqD:2}, are identically satisfied. On the other hand, from the mixed equations proportional to $\Delta_a\left(\ell\right)$, i.e., Eqs. \eqref{eqD:3} and \eqref{eqD:4}, one must have  $\bar\mu=\bar Q=0$ for spacelike hypersurfaces,  $\bar Q=\bar p +\bar \Pi=0$ for timelike hypersurfaces, and  $\bar\mu=\bar p + \bar \Pi=\pm\bar Q$ for null hypersurfaces. Notably, the conditions on the thermodynamical quantities $\bar\mu$, $\bar p$, $\bar \Pi$, and $\bar Q$ also follow directly from Eqs. \eqref{aQ} to \eqref{JuncFin}, although such a result is not obvious due to the larger complexity of the equations, which implies that Eqs. \eqref{eqD:1} to \eqref{eqD:4} provide redundant information.
 
The type II junction conditions we have derived show that junction conditions can be (and maybe ought to be) seen in a new perspective. By expressing the junction conditions in terms of the 1+1+2 potentials, we have, in fact, obtained conditions for which two specific congruences in two different spacetimes can be joined to be continuous across the boundary. As these congruences are associated with specific observers in the covariant formalism, the equations above reveal that junction conditions are generally only valid for a particular class of observers.  In the standard formulation of junction conditions, the choice of an observer is intrinsic to the choice of the coordinate system in the two spacetimes, and this might lead to the idea that, upon using the junction conditions, one simply matches two geometries. Our results imply that the underlying idea of the junction conditions is that one matches geometries as seen by two specific observers.  

\subsection{Propagation of constraints}

Because of the structure of the 1+1+2 scalars, the algebraic constraints in Eqs.~\eqref{heqs}, \eqref{gauscurv}, and Eq.~\eqref{Other_constr}
seem not to contribute to the junction conditions. Nevertheless, the dot and hat derivative of such constraints might still contain some relevant conditions.

It turns out that all the derivatives of the constraints Eq.~\eqref{gauscurv} and Eq.~\eqref{Other_constr} lead to an identity, upon using the $1+1+2$ equations to eliminate the jumps of the quantities that appear upon differentiating. However, taking the covariant derivative of Eq.\eqref{heqs}, expanding these derivatives in terms of the {\it dot} and {\it hat} derivatives, using Eqs.\eqref{dotdist} and \eqref{hatdist} to write the results in terms of the jumps of the potentials, and keeping only the singular terms proportional to $\delta\left(\ell\right)$ (the singular terms proportional to $\Delta_a\left(\ell\right)$ vanish identically due to Eq. \eqref{solweylH}), one obtains
\begin{eqnarray}
&&\left(\tau u^a+\varsigma e^a\right)\left[3\langle\xi\rangle\left[\Sigma\right]_{\pm}+3\langle\Sigma\rangle\left[\xi\right]_{\pm}-\left[\mathcal H\right]_{\pm}-\right.\nonumber \\
&&\left.-\left(2\langle\mathcal A\rangle-\langle\phi\rangle\right)\left[\Omega\right]_{\pm}+\langle\Omega\rangle\left(2\left[\mathcal A\right]_{\pm}-\left[\phi\right]_{\pm}\right)\right]=0.\label{propconst}
\end{eqnarray}
This equation must be satisfied for any $\tau$ and $\varsigma$. If $\tau=1$, which implies that $\varsigma=0$, one can use the evolution equations from Eqs.\eqref{aQ} to \eqref{aomega} to replace the jumps of the geometrical quantities and obtain the additional constraint
\begin{equation}\label{timeconst}
\left(2\left[\theta\right]_{\pm}+3\bar p-\bar \mu\right)\langle\xi\rangle-2\langle\Omega\rangle\left[\mathcal A\right]_{\pm}=0.
\end{equation}
On the other hand, for the case of $\varsigma=1$, and consequently $\tau=0$, one uses instead the propagation equations from Eqs.\eqref{propphi} to \eqref{bomega}, as well as the mixed equation in Eq.\eqref{propa}, to cancel the jumps of the geometrical quantities and obtain the additional constraint
\begin{equation}\label{spaceconst}
\left(\left[\theta\right]_{\pm}-\bar Q\right)\langle\xi\rangle-\left(\bar p + \bar \Pi\right)\langle\Omega\rangle=0.
\end{equation}
Finally, for the cases of outgoing and ingoing null hypersurfaces with $\tau=1/\sqrt{2}$ and $\varsigma=\pm 1/\sqrt{2}$, both of the projections of Eq.\eqref{propconst} in $u^a$ and $e^a$ are independently non-zero. A combination of the two procedures described above for $\tau=1$ and $\varsigma=1$ has to be followed. In this case, each of the projections contributes with an additional constraint, which are
\begin{equation}\label{nullconst1}
\left(2\left[\theta\right]_{\pm}+\sqrt{2}\left(3\bar p + \bar \mu\right)\right)\left(\langle\xi\rangle\mp\langle\Omega\rangle\right)=0,
\end{equation}
\begin{equation}\label{nullconst2}
\left(\left[\theta\right]_{\pm}\mp\sqrt{2}\bar Q\right)\langle\xi\rangle\mp\left(\left[\theta\right]_{\pm}+\sqrt{2}\left(\bar p+\bar\Pi\right)\right)\langle\Omega\rangle=0.
\end{equation}
where the upper and lower signs correspond to outgoing and ingoing null hypersurfaces, respectively.

For LRS-II spacetimes, these constraints do not influence the junction conditions since these spacetimes have $\xi=\Omega=0$ and thus Eqs.\eqref{timeconst} to \eqref{nullconst2} are automatically satisfied. However, for  LRS-I and LRS-III spacetimes, for which $\Omega\neq 0$ and $\xi\neq 0$ respectively, these equations might constrain the junction conditions or even contribute with additional junction conditions to the system, and thus they must be taken into account simultaneously with the singular equations obtained in the previous subsection.

\section{Properties of junctions of LRS spacetimes}\label{sec:junctions}
In this section, we use the formalism and equations derived in the previous section to deduce some general prescriptions on the junction conditions in LRS spacetimes. 
\subsection{Comoving observers}
We start with junction conditions valid for comoving LRS observers. As a first observation, we note that some junction conditions can be derived independently of the type of hypersurface considered. From Eqs.\eqref{bxi} and \eqref{axi}, one verifies that, independently of the value of the parameters $\tau$ and $\varsigma$, the jump $\left[\xi\right]_{\pm}$ is always forced to vanish. The same applies to the jump $\left[\Omega\right]_{\pm}$ if one considers instead Eqs.\eqref{bomega} and \eqref{aomega}. One thus obtains
\begin{equation}\label{junctxi}
\left[\xi\right]_{\pm}=0,
\end{equation}
\begin{equation}\label{junctomega}
\left[\Omega\right]_{\pm}=0.
\end{equation}
i.e. (comoving observers in an) LRS spacetime can be joined only if they have zero vorticity or twist or if these quantities match at the boundary. However,  the 1+1+2 equations indicate that if those two quantities are zero in an event, they must be zero everywhere. Consequently, for comoving LRS observers, LRS-II spacetimes cannot be matched to LRS-I or LRS-III spacetimes.

The remaining junction conditions for different hypersurfaces can now be extracted from these equations by the selection of a normal vector, i.e., by choosing the values of $\tau$ and $\varsigma$ among the choices available in Eq. \eqref{sigtauchoice}. We shall consider separately the cases of LRS-I, LRS-II, and LRS-III spacetimes, excluding the combinations between normal, vorticity, and twist mentioned at the end of Sec. \ref{sec:1ParForm}.  The junction conditions obtained by the resolution of the system of Eqs.\eqref{bQ} to \eqref{JuncFin} are summarized in Tables \ref{tab:timelike}, \ref{tab:spacelike} and \ref{tab:null}.  

The first noteworthy result from the Tables above is that, depending on the type of boundary surface, there are different key variables characterizing the junctions. For example, for timelike surfaces, an important role is played by the jump of the expansion and the shear\footnote{See also \cite{Fayos:1996gw} for the role of the expansion in junction conditions in coordinates.}, while for spacelike surfaces, the prominent role is played by the jump of the acceleration and the expansion of the spacelike congruence. This is consistent with the idea that the type II junction conditions are connected to the jump of the extrinsic curvature. We explicitly connect those quantities with the extrinsic curvature in Section \ref{ExtCurv112} of Appendix \ref{Null_K}.

Another interesting feature concerns timelike hypersurface (Table \ref{tab:timelike}). In such a case, our results indicate that a thin shell presents a zero energy density but non-zero pressure terms. This result shows that in joining, e.g., expanding cosmologies (or an overdensity of matter in a cosmological background), a special surface between the two geometries exists, generated by the difference in the expansion rate and shear of the two spacetimes. This surface cannot be interpreted as a proper matter shell but rather as a ``frontier''  which, at least in specific cases, could be associated with the concept of ``turnaround radius'' in non-homogeneous cosmologies (see, e.g., \cite{Mimoso:2009wj}).  Indeed, the existence of turnaround surfaces was shown to exist in the context of McVittie spacetimes \cite{Nandra:2011ug,Nandra:2011ui,Kaloper:2010ec}, one of the few examples of analytical inhomogeneous spacetimes. In addition, the component of the pressure orthogonal to the surface is not zero, suggesting this type of boundary should have some form of dynamics.

In the case of spacelike surfaces, instead, the presence of a thin shell is controlled by the jump of the variables $\phi$ and $\mathcal A$. It is clear that the radial pressure of the thin shell is always zero and that, therefore, such shells are not dynamic objects. Indeed, differently from the case of timelike surfaces, they can be represented by a matter distribution. These are the typical shells that describe compact stars or more exotic objects like gravastars. 

In the case of a null boundary, one observes, as expected, a mixture of the conditions of the two cases above. The additional element is a non-trivial conservation law originating from the tensor $S_{ab}$, which, in turn, points towards a more complex behavior of matter on the shell. In addition, in this case, the junction conditions related to Bianchi identities lead to two differential equations describing the matter conservation on the shell. 

Another interesting result to highlight concerns the value of the magnetic part of the Weyl tensor. In the case of matching between LRS-II spacetimes, this quantity is always identically zero, but it does not have to be so in other LRS spacetimes. Our analysis shows that within LRS-I and LRS-III spacetimes, one could have a matching between spacetimes with different values of the magnetic part of the Weyl tensor. In both cases, this match can only be obtained by introducing a thin shell. 
The relations we found relate the jump in  $\mathcal H$ to the vorticity/twist of the spacetime. As normally $\mathcal H$ is connected with the presence of gravitational radiation, the junction conditions we derive point out to the possibility that there might exist types of thin shells able to ``absorb'' gravitational radiation, in the sense of encompassing a hypervolume of spacetime with less (or no) gravitational radiation. 

\begin{table*}
    \centering
    \begin{tabular}{c|c|c|c|c}
        & LRS-II & LRS-III \\ \hline
        $\bar\mu$& 0 & 0  \\
        $\bar p$ &  $-\frac{2}{3}\left[\theta\right]_{\pm}$ & $-\frac{2}{3}\left[\theta\right]_{\pm}$\\
        $\bar \Pi$ &  $\left[\Sigma\right]_{\pm}$ & $\left[\Sigma\right]_{\pm}$   \\
        $\bar Q$ & 0 & 0 \\
        $\left[\phi\right]_{\pm}$ & 0 & 0 \\
        $\left[\mathcal A\right]_{\pm}$  & ind. & ind. \\
        $\left[\theta\right]_{\pm}$ & ind. & ind. \\
        $\left[\Sigma\right]_{\pm}$& ind. & ind.\\
        $\left[\mathcal E\right]_{\pm}$  & $-\frac{1}{2}\left[\Pi\right]_{\pm}-\left[\Sigma\right]_{\pm}\left(\langle\Sigma\rangle-\frac{1}{3}\langle\theta\rangle\right)+\frac{1}{3}\langle\Sigma\rangle\left[\theta\right]_{\pm}$ & $-\frac{1}{2}\left[\Pi\right]_{\pm}-\left[\Sigma\right]_{\pm}\left(\langle\Sigma\rangle-\frac{1}{3}\langle\theta\rangle\right)+\frac{1}{3}\langle\Sigma\rangle\left[\theta\right]_{\pm}$ \\
        $\left[\mathcal H\right]_{\pm}$& 0 & $3\langle\xi\rangle\left[\Sigma\right]_{\pm}$ \\
        $\left[\mu\right]_{\pm}$  & $\frac{2}{3}\langle\theta\rangle\left[\theta\right]_{\pm}-\frac{3}{2}\langle\Sigma\rangle\left[\Sigma\right]_{\pm}$ &  $\frac{2}{3}\langle\theta\rangle\left[\theta\right]_{\pm}-\frac{3}{2}\langle\Sigma\rangle\left[\Sigma\right]_{\pm}$ \\
        $\left[p\right]_{\pm}$ & ind. & ind. \\
        $\left[\Pi\right]_{\pm}$  & ind. & ind. \\
        $\left[Q\right]_{\pm}$ & $\frac{2}{3}\langle \mathcal A \rangle\left[\theta\right]_{\pm}-\left[\Sigma\right]_{\pm}\left(\langle \mathcal A \rangle+\frac{3}{2}\langle \phi \rangle\right)$ & $\langle \mathcal A \rangle\left(\frac{2}{3}\left[\theta\right]_{\pm}-\left[\Sigma\right]_{\pm}\right)$\\
    \end{tabular}
    \caption{Junction conditions for matching two spacetimes along a spacelike hypersurface for LRS spacetimes. The entry "ind." indicates that the quantity is an independent parameter.}
    \label{tab:timelike}
\end{table*}

\begin{table*}
    \centering
    \begin{tabular}{c|c|c|c|c}
        &  LRS-I & LRS-II  \\ \hline
        $\bar\mu$ &  $-\left[\phi\right]_{\pm}$ & $-\left[\phi\right]_{\pm}$ \\
        $\bar p$ &  $\frac{1}{3}\left[\phi\right]_{\pm}+\frac{2}{3}\left[\mathcal A\right]_{\pm}$ & $\frac{1}{3}\left[\phi\right]_{\pm}+\frac{2}{3}\left[\mathcal A\right]_{\pm}$ \\
        $\bar \Pi$ & $-\frac{1}{3}\left[\phi\right]_{\pm}-\frac{2}{3}\left[\mathcal A\right]_{\pm}$ & $-\frac{1}{3}\left[\phi\right]_{\pm}-\frac{2}{3}\left[\mathcal A\right]_{\pm}$\\
        $\bar Q$ &  0 & 0 \\
        $\left[\phi\right]_{\pm}$ &  ind. & ind.  \\
        $\left[\mathcal A\right]_{\pm}$ &  ind. & ind.  \\
        $\left[\theta\right]_{\pm}$ &  0 & ind. \\
        $\left[\Sigma\right]_{\pm}$ &  0 & $\frac{2}{3}\left[\theta\right]_{\pm}$ \\
        $\left[\mathcal E\right]_{\pm}$ &  $\frac{1}{3}\left[\mu\right]_{\pm}-\frac{1}{2}\left[\Pi\right]_{\pm}+\frac{1}{2}\langle \phi \rangle\left[\phi\right]_{\pm}$ & $\frac{1}{3}\left[\mu\right]_{\pm}-\frac{1}{2}\left[\Pi\right]_{\pm}+\frac{1}{2}\langle \phi \rangle\left[\phi\right]_{\pm}$  \\
        $\left[\mathcal H\right]_{\pm}$ & $\langle \Omega \rangle\left(\left[\phi\right]_{\pm}-2\left[\mathcal A\right]_{\pm}\right)$ & 0 \\
        $\left[\mu\right]_{\pm}$ &  ind. & ind.  \\
        $\left[p\right]_{\pm}$ & $-\left[\Pi\right]_{\pm}+\left[\phi\right]_{\pm}\left(\langle \mathcal A \rangle+\frac{1}{2}\langle \phi \rangle\right)+\langle \phi \rangle\left[\mathcal A\right]_{\pm}$ & $-\left[\Pi\right]_{\pm}+\left[\phi\right]_{\pm}\left(\langle \mathcal A \rangle+\frac{1}{2}\langle \phi \rangle\right)+\langle \phi \rangle\left[\mathcal A\right]_{\pm}$ \\
        $\left[\Pi\right]_{\pm}$ & ind. & ind. \\
        $\left[Q\right]_{\pm}$ &  0 & $\frac{1}{6}\left[\phi\right]_{\pm}\left(4\langle\theta\rangle+3\langle\Sigma\rangle\right)+\left[\mathcal A\right]_{\pm}\left(\langle \Sigma \rangle-\frac{2}{3}\langle\theta\rangle\right)$ \\
    \end{tabular}
    \caption{Junction conditions for matching two spacetimes along a timelike hypersurface for LRS spacetimes. The entry "ind." indicates that the quantity is an independent parameter.}
    \label{tab:spacelike}
\end{table*}

\begin{table*}
    \centering
    \begin{tabular}{c|c|c|}
& LRS-II \\ \hline
$\bar\mu$ & $\mp \frac{\sqrt{2}}{2}\left[\phi\right]_{\pm}$  \\
$\bar p$ & $\pm \frac{\sqrt{2}}{6}\left[\phi\right]_{\pm}\pm\frac{\sqrt{2}}{3}\left[\mathcal A\right]_{\pm}-\frac{\sqrt{2}}{3}\left[\theta\right]_{\pm}$\\
$\bar\Pi$ & $\mp\frac{2\sqrt{2}}{3}\left[\phi\right]_{\pm}\mp\frac{\sqrt{2}}{3}\left[\mathcal A\right]_{\pm}+\frac{\sqrt{2}}{3}\left[\theta\right]_{\pm}$  \\
$\bar Q$ &  $\frac{\sqrt{2}}{2}\left[\phi\right]_{\pm}$  \\
$\left[\phi\right]_{\pm}$  & ind. \\
$\left[\mathcal A\right]_{\pm}$  & ind.  \\
$\left[\theta\right]_{\pm}$  & ind. \\
$\left[\Sigma\right]_{\pm}$  & $\frac{2}{3}\left[\theta\right]_{\pm}\mp\left[\phi\right]_{\pm}$  \\
$\left[\mathcal E\right]_{\pm}$ &  $\frac{1}{3}\left[\mu\right]_{\pm}-\frac{1}{2}\left[\Pi\right]_{\pm}+\frac{1}{2}\langle \phi \rangle\left[\phi\right]_{\pm}\pm\left[\phi\right]_{\pm}\left(\frac{1}{2}\langle\Sigma\rangle-\frac{1}{3}\langle\theta\rangle\right)$  \\
$\left[\mathcal H\right]_{\pm}$  & 0  \\
$\left[\mu\right]_{\pm}$  & ind.  \\
$\left[p\right]_{\pm}$ &  $\left[\mu\right]_{\pm}-\left[\Pi\right]_{\pm}+\left[\phi\right]_{\pm}\left(3\langle \phi \rangle+4\langle \mathcal A \rangle\right)\mp 2\left[\phi\right]_{\pm}\left(\frac{3}{2}\langle\Sigma\rangle+\langle\theta\rangle\right)+\left(\left[\theta\right]_{\pm}\mp\left[\mathcal A\right]_{\pm}\right)\left(\langle\Sigma\rangle-\frac{2}{3}\langle\theta\rangle\mp\langle \phi \rangle\right)+\dot{\bar\mu}-\left(\hat{\bar p}+\hat{\bar\Pi}\right)\mp\dot{\bar Q}\pm\hat{\bar Q}$ \\
$\left[\Pi\right]_{\pm}$  & ind. \\
$\left[ Q\right]_{\pm}$ &  $\mp\left[\mu\right]_{\pm}\mp\left[\phi\right]_{\pm}\left(\langle \phi \rangle+2\langle \mathcal A \rangle\right)+2\left[\phi\right]_{\pm}\left(\langle\Sigma\rangle+\frac{1}{3}\langle\theta\rangle\right)+\left(\left[\mathcal A\right]_{\pm}\mp\left[\theta\right]_{\pm}\right)\left(\langle\Sigma\rangle-\frac{2}{3}\langle\theta\rangle\right)-\hat{\bar Q}\mp\dot{\bar\mu}$ \\
    \end{tabular}
    \caption{Junction conditions for matching two spacetimes along outgoing and ingoing null hypersurface for LRS spacetimes. In the case of multiple signs, the upper sign corresponds to outgoing hypersurfaces, and the lower sign corresponds to ingoing hypersurfaces. The entry "ind." indicates that the quantity is an independent parameter.}
    \label{tab:null}
\end{table*}

\subsection{Tilted observers}\label{sec:exs}
The relations derived in the previous section concern only observers comoving with matter in the spacetimes to be matched. However, limiting the junction conditions to these cases would be reductive as, in general, one can also consider non-comoving (tilted) observers in a given spacetime.  In this section, we extend our reasoning to the case of such observers. 

In the 1+1+2 formalism, tilted observers can be characterized by the following timelike and spacelike vectors
\begin{equation}
\begin{split}\label{boost_u_e}
\breve{u}^a&= {u}^a \cosh \beta + {e}^a \sinh \beta\\
\breve{e}^a&= {e}^a \cosh \beta + {u}^a \sinh \beta
\end{split}
\end{equation}
where the angle $\beta$ is defined by $\cosh \beta=-\breve{u}^a{u}_a$. The relations \eqref{boost_u_e} are nothing but the Lorentzian boost of the vectors ${u}^a $ and ${e}^a$. Under these transformations, the kinematic 1+1+2 potentials transform as
\begin{equation}
\begin{split}
\breve{\phi}=&\phi \cosh \beta + \left(\frac{2}{3}\Theta+ \Sigma\right)\sinh \beta \\
\breve{\mathcal A} =&{\mathcal A} \cosh \beta+ \left(\frac{1}{3}\Theta+ \Sigma\right)\sinh \beta \\
&+ ({u}^a \cosh \beta + {e}^a \sinh \beta)\nabla_a \beta\\
\breve{\Theta}=& \Theta \cosh \beta + (\phi+{\mathcal A} ) \sinh \beta \\
&+ ({e}^a \cosh \beta + {u}^a \sinh \beta) \nabla_a \beta\\
\breve{\Sigma}=& \Sigma \cosh \beta- \frac{1}{3} (\phi-2{\mathcal A}) \sinh \beta \\
&+ \frac{2}{3} ({e}^a \cosh \beta + {u}^a \sinh \beta) \nabla_a \beta\\
\breve{\Omega}=& -4 (\Omega \cosh \beta + \xi \sinh \beta) \\
\breve{\xi}=& -4(\xi \cosh \beta + \Omega \sinh \beta) \\
\breve{\mathcal E}=& \mathcal E \\
\breve{\mathcal H}=& \mathcal H \\
\end{split}
\end{equation}
whereas the thermodynamical variables  transform as
\begin{equation}
\begin{split}
\breve{\mu}=&\mu +\left(\mu+p+\Pi\right)\sinh^2 \beta -Q \sinh (2\beta) \\
\breve{p}=&p +\frac{1}{3}\left(\mu+p+\Pi\right)\sinh^2 \beta -\frac{1}{3} Q \sinh (2\beta) \\
\breve{Q}=&Q \cosh(2\beta) -\left(\mu+p+\Pi\right)\sinh \beta\cosh \beta  \\
\breve{\Pi}=&\Pi +\frac{2}{3}\left(\mu+p+\Pi\right)\sinh^2 \beta-\frac{2}{3}Q \sinh (2\beta)   
\end{split}
\end{equation}
At this point, one can use the junction conditions listed in Tables \ref{tab:timelike}, \ref{tab:spacelike}, \ref{tab:null}, in which the transformed quantities are used instead of the conventional ones. For later convenience, we introduce the ``tilted jump'' $\left\{X\right\}_{\pm}$ of a given scalar $X$ as
\begin{equation}
\left\{X\right\}_{\pm}=\breve{X}^+_H-X^-_H
\end{equation}
which is used to characterize the modified type II junction conditions. For simplicity's sake, we refer to these conditions as ``tilted type II junction conditions''.We provide an example of the application of these conditions in the following sections.

Some considerations on the general form of the transformations are in order. First, we should notice that a fluid that is perfect in its rest frame is not necessarily perfect for non-comoving observers. This is, indeed, a well-known result, and we refer the reader to the relevant literature on the matter \cite{King:1972td}. Second, the boost leaves the scalars associated with the Weyl tensor unchanged; therefore, the conclusions on the ``absorbing'' thin shells mentioned in the previous section also remain valid in this case.  Third, the transformation of the vorticity and the twist shows that LRS-I or LRS-III spacetimes can be matched to other spacetimes only if the congruence of the observer in the exterior spacetime has vorticity or twist. In other words, no LRS-II spacetime can be joined with LRS-I or LRS-III geometries.

\section{Examples}\label{sec:exs}
In this section, we apply the junction conditions we derived in the previous sections to some physically relevant cases. As mentioned before, the covariant junction conditions amount to the search of two congruences, one in the interior and one in the exterior spacetimes, which can be made to match at the boundary. The procedure we adopt is the following. First, we write the line elements on the interior and exterior spacetimes in the coordinate system of choice and select the type of boundary and its normal vector. Then, we select a working coordinate system, which can be one of the coordinates chosen for the two spacetimes or a different one altogether. Next, using the normal vector, we fix the congruences $u_a$ and $e_a$ in the first spacetime. The first type I junction condition in Eq. \eqref{n-n} can then be used to determine the corresponding normal vector in the other spacetime and, therefore, the corresponding vectors $u_a$ and $e_a$. Then, the second type I junction condition in Eq. \eqref{JC_1} determines the relation between the two coordinate systems at the boundary. At this point, one can calculate the 1+1+2  quantities necessary to write the non-trivial type II junction conditions from Eqs.~\eqref{bQ} to \eqref{JuncFin} directly or using the formulas in Appendix \ref{1+1+2 coord}. Depending on the relation between the comoving LRS observer congruence in the second spacetime and the congruence determined by the normal of the first spacetime, one can choose to use tilted type II junction conditions. Applying type I and II junction conditions entirely determines the geodesic congruence in the second spacetime in the working coordinates.  

\subsection{The Martinez thin-shell}

\subsubsection{Spherical coordinates}

As a first example, let us look into the Martinez thin-shell in spherical coordinates \cite{Martinez:1989hn,Brady:1991np}. This is a solution in general relativity of a thin shell separating an interior Minkowski spacetime from an exterior Schwarzschild spacetime. Both solutions are vacuum solutions, i.e., $T_{ab}^+=T_{ab}^-=0$ and therefore $\left[\mu\right]_{\pm}=0$, $\left[p\right]_{\pm}=0$,  $\left[\Pi\right]_{\pm}=0$. The line elements describing the interior and exterior spacetimes of the Martinez shell are of the forms
\begin{equation}\label{martmink}
ds_-^2=-d {t}^2+d{r}^2+{r}^2d{\Omega}^2,
\end{equation}
\begin{equation}\label{martschw}
ds_+^2=-\left(1-\frac{M}{4\pi \tilde r}\right)d\tilde t^2+\left(1-\frac{M}{4\pi \tilde r}\right)^{-1}d\tilde r^2+r^2d\tilde \Omega^2,
\end{equation}
where
\begin{equation}
\begin{split}
d\Omega^2&=d\theta^2+\sin^2\theta d\phi^2\\
d\tilde\Omega^2&=d\tilde\theta^2+\sin^2\theta d\tilde\phi^2
\end{split}
\end{equation}
 are the 2-sphere line elements. We choose the coordinates of the interior Minkowski spacetime as a working coordinate system.  Consider a spacelike boundary characterized by the equation $\tilde{r}= const.=\tilde{r}_H$. Then, the (spacelike) normal vector in the interior spacetime can be chosen as
\begin{equation}
n^-_a d\tilde{x}^a=d\tilde{r}.
\end{equation}
which is spacelike. Then, selecting $e_a$ parallel to $n_a$  one can write
\begin{equation}
e_a^-d\tilde{x}^a=n^-_a d\tilde{x}^a=d\tilde{r},
\end{equation}
since by definition $e_a^-u_a^-g^{ab}=0$ and assuming $u_a^-u_a^-g^{ab}=-1$ we have
\begin{equation}
u_a^- d\tilde{x}^a=-d\tilde{t}.
\end{equation}
so that the congruence in the interior spacetime represents a comoving LRS observer.
The first type I junction condition in Eq. \eqref{n-n} implies that\footnote{Notice that we have assumed here that the spatial coordinates of the exterior and exterior spacetime do not mix. The form of the two metrics suggests this assumption. One can redo the calculations considering a more general relation between the coordinates. The second type I junction condition then provides prescriptions on the coordinate dependence necessary to perform the junction. We show an example of this procedure in the case of the Oppenheimer-Snyder collapse.}
\begin{equation}
e_a^+dx^a=\frac{\partial\tilde{r}}{\partial r}e_1^-dr=\left(1-\frac{M}{4 \pi r}\right)^{-1/2} d r,
\end{equation}
and, following the same procedure to obtain $u_a^-$, we arrive at
\begin{equation}
u_a^+ dx^a=\frac{\partial\tilde{t}}{\partial t}u_0^-dt=-\left(1-\frac{M}{4 \pi r}\right)^{1/2}  d t.
\end{equation}
The congruence in the exterior spacetime represents a comoving LRS observer, so tilted junction conditions are unnecessary. The tensor $q_{ab}$ in $H$ for  the two spacetimes can be written in terms of line elements as 
\begin{equation}\label{martmink}
ds_{H,-}^2= q_{ab}^- d{x}^a d {x}^b=-d {t}^2+{r}^2_Hd{\Omega}^2,
\end{equation}
\begin{equation}\label{martschw}
\begin{split}
ds_{H,+}^2&= q_{ab}^+ d\tilde x^a d\tilde x^b=\\
&=-\left(1-\frac{M}{4 \pi \tilde r_H}\right)d\tilde t^2+\tilde r^2_Hd\tilde\Omega^2=\\
&=-d {t}^2+r^2_Hd{\Omega}^2,
\end{split}
\end{equation}
where in the last step we have  written $ds_{H,+}^2$ in the working coordinate system assuming $d\tilde{\Omega}_H^2=d\Omega_H^2$ i.e. we assume that the angular coordinates coincide on $H$. The second type I junction condition, Eq. \eqref{JC_1}, implies 
\begin{equation}
\tilde{r}_H=r_H.
\end{equation}
As the congruences associated with $u_a^+$, $e_a^+$, $u_a^-$ and $e_a^-$ are all rotation free $\Omega=0=\xi$ and also have $\theta=0$ and $\Sigma=0$, the only non-trivial conditions are
\begin{equation}\label{jcLRS-Ii}
\begin{split}
\bar\mu&=-\left[\phi\right]_{\pm}, \\
\bar p &=\frac{1}{3}\left[\phi\right]_{\pm}+\frac{2}{3}\left[\mathcal A\right]_{\pm}, \\
\bar\Pi &=-\frac{1}{3}\left[\phi\right]_{\pm}-\frac{2}{3}\left[\mathcal A\right]_{\pm}, \\
\left[\mathcal E\right]_{\pm}&=\frac{1}{2}\langle \phi \rangle\left[\phi\right]_{\pm}, \\
\left[p\right]_{\pm}&=\left[\phi\right]_{\pm}\left(\langle \mathcal A \rangle+\frac{1}{2}\langle \phi \rangle\right)+\langle \phi \rangle\left[\mathcal A\right]_{\pm}.
\end{split}
\end{equation}
At $r=r_H$ one obtains
\begin{equation}\label{spherJphi}
\left[\phi\right]_{\pm}=-\frac{2}{r_H}\left(1-\sqrt{1-\frac{M}{4 \pi r_H}}\right),
\end{equation}
\begin{equation}
\left[\mathcal A\right]_{\pm}=\frac{M}{r_H^2}\left(1-\frac{M}{4 \pi r_H}\right)^{-\frac{1}{2}}.
\end{equation}
The first three conditions in Eq. \eqref{jcLRS-Ii} then  show the presence of a thin-shell at $r_H$ characterized by a stress-energy tensor $S_{ab}$ in Eq. \eqref{Sab} with
\begin{equation}
\bar\mu=\frac{2}{r_H}\left(1-\sqrt{1-\frac{M}{4 \pi r_H}}\right),
\end{equation}
\begin{equation}
\bar p_e
=\frac{1}{r_H}\left[\left(1-\frac{M}{r_H}\right)\left(1-\frac{M}{4 \pi r_H}\right)^{-\frac{1}{2}}-1\right],
\end{equation}
\begin{equation}
\bar p_N =
0,
\end{equation}
where $p_e$ is the radial pressure and $p_N$ is the transverse pressure. 

Of the last two conditions in Eq. \eqref{jcLRS-Ii}, the one for $\left[p\right]_{\pm}$ is automatically satisfied by Eq. \eqref{spherJphi}, and the one for $\left[\mathcal E\right]_{\pm}$ can be verified immediately by direct calculation. In fact,
\begin{equation}\label{martE}
\left[\mathcal E\right]_{\pm}=-\frac{M}{4 \pi r_H^3}.
\end{equation}
and the $\langle \phi \rangle$ appearing on the L.H.S. of  the second last of  Eq. \eqref{jcLRS-Ii}, can be written, in $H$, as
\begin{equation}
\langle \phi \rangle=-\frac{1}{r_H}\left(1+\sqrt{1-\frac{M}{4 \pi r_H}}\right),
\end{equation}
therefore
\begin{equation}
\langle \phi \rangle[\phi]_{\pm}=\frac{M}{2 \pi r_H^3}.
\end{equation}
This result is also obtained if one considers that for static, spherically symmetric and vacuum spacetimes, like Eq.~\eqref{martmink} and Eq.~\eqref{martschw}, Eq.~\eqref{lusimthe} implies $\mathcal E=- \mathcal A \phi$.

\subsection{The Schwarzschild fluid star}

Let us now consider the case of the Schwarzschild fluid star. This solution of Einstein's field equations consists of an interior solution with a constant density perfect fluid \cite{schwarzschildstar} matched to an exterior vacuum Schwarzschild solution.

The line elements for the interior and exterior spacetimes are given respectively in the usual spherical coordinates by
\begin{eqnarray}
ds^2_-&=&-\frac{1}{4}\left(3\sqrt{1-\frac{M}{4\pi R}}-\sqrt{1-\frac{M \tilde{r}^2}{4\pi R^3}}\right)^2d\tilde{t}^2+\nonumber \\
&+&\left(1-\frac{M \tilde{r}^2}{4\pi R^3}\right)^{-1}d\tilde{r}^2+\tilde{r}^2d\tilde{\Omega}^2,\label{intschw}
\end{eqnarray}
\begin{equation}\label{extschw}
ds_+^2=-\left(1-\frac{M}{4 \pi r}\right)dt^2+\left(1-\frac{M}{4 \pi r}\right)^{-1}dr^2+r^2d\Omega^2,
\end{equation}
where $M$ is the total mass of the star, and $R$ is the radius of the star. 

The interior solution is non-vacuum, while the stress-energy tensor $T_{ab}$ vanishes for the exterior solution. The interior source is described by an isotropic perfect fluid, i.e., 
\begin{equation}
(T^-)_a^b=\text{diag}\left(-\mu^-_0,p^-,p^-,p^-\right),
\end{equation}
where $\mu_0$ is the constant density of the fluid and $p^-=p^-\left(r\right)$ is the isotropic pressure
\begin{equation}\label{PressSchw}
p^-\left(r\right)=\mu_0\frac{\sqrt{1-\frac{Mr^2}{4\pi R^3}}-\sqrt{1-\frac{M}{4 \pi r}}}{3\sqrt{1-\frac{M}{4 \pi r}}-\sqrt{1-\frac{Mr^2}{4\pi R^3}}},
\end{equation}
which corresponds to the well-known solution of the TOV equations. Since the fluid considered is isotropic, this result implies that $\Pi^-=0$ and $p^-_e=p^-_N=p^-$. 

We choose the coordinate system of the interior solution as the working coordinate system. If we consider a spacelike boundary characterized by the equation $\tilde{r}=\tilde{r}_H=const.$, the (spacelike) normal vector in the interior spacetime  can be chosen as
\begin{equation}
n^-_a d\tilde{x}^a=\left(1-\frac{M\tilde{r}^2}{4\pi R^3}\right)^{-1/2}d\tilde{r}.
\end{equation}
which is spacelike. Choosing $e^-_a$ parallel to $n^-_a$ we have
\begin{equation}
e_a^-d\tilde{x}^a=n^-_a d\tilde{x}^a=\left(1-\frac{M \tilde{r}^2}{4\pi R^3}\right)^{-1/2}d\tilde{r},
\end{equation}
and therefore $u^-_a$ is
\begin{equation}
u_a^- d\tilde{x}^a=-\frac{1}{2}\left|3\sqrt{1-\frac{M}{4\pi R}}-\sqrt{1-\frac{M \tilde{r}^2}{4\pi R^3}}\right|d\tilde{t},
\end{equation}
so that the congruence of the interior spacetime represents a comoving LRS observer. Equation \eqref{n-n} implies that 
\begin{equation}
e_a^+dx^a=\frac{\partial\tilde{r}}{\partial r}e_1^- dr=\left(1-\frac{M}{4 \pi r}\right)^{-1/2} d r,
\end{equation}
and choosing as usual $u_a^+$ normalized and orthogonal to $e^+_a$ we have 
\begin{equation}
u_a^+ d\tilde{x}^a=\frac{\partial\tilde{t}}{\partial t}u_1^- dt=-\left(1-\frac{M}{4 \pi r}\right)^{1/2} dt
\end{equation}
and also the congruence of the exterior spacetime represents a comoving LRS observer. The tensor $q_{ab}$ in $H$ for  the two spacetimes can be written in terms of line elements as
\begin{equation}
\begin{split}
ds_{H,-}^2=&-\frac{1}{4}\left(3\sqrt{1-\frac{M}{4\pi R}}-\sqrt{1-\frac{M\tilde{r}_H^2}{4\pi R^3}}\right)^2d\tilde{t}^2+\\
&+\tilde{r}_H^2d\tilde{\Omega}^2,
\end{split}
\end{equation}
\begin{equation}
\begin{split}
ds_{H,+}^2=&-\left(1-\frac{M}{4 \pi r_H}\right)dt^2+r_H^2d\Omega^2\\
=&-\frac{1}{4}\left(3\sqrt{1-\frac{M}{4\pi R}}-\sqrt{1-\frac{M\tilde{r}_H^2}{4\pi R^3}}\right)^2d\tilde{t}^2+\\
&+{r}_H^2d{\Omega}^2,
\end{split}
\end{equation}
which, once written in the working coordinates as in the previous example, gives  $r_H=\tilde{r}_H$ if $d\tilde{\Omega}_H^2=d\Omega_H^2$.

We are now ready to look at the type II junction conditions. As the $e_a^+$ and $u_a^+$ are associated with a comoving LRS observer, tilted junction conditions are unnecessary. In addition, since the congruences $u_a^+$, $e_a^+$, $u_a^-$, and $e_a^-$ are rotation-free and have zero expansion and shear, one obtains the following non-trivial type II junction conditions
\begin{equation}\label{jcLRS-Iischw}.
\begin{split}
\bar\mu&=-\left[\phi\right]_{\pm}, \\
\bar p &=\frac{1}{3}\left[\phi\right]_{\pm}+\frac{2}{3}\left[\mathcal A\right]_{\pm}, \\
\bar\Pi &=-\frac{1}{3}\left[\phi\right]_{\pm}-\frac{2}{3}\left[\mathcal A\right]_{\pm}, \\
\left[\mathcal E\right]_{\pm}&=\frac{1}{3}\left[\mu\right]_{\pm}+\frac{1}{2}\langle \phi \rangle\left[\phi\right]_{\pm}, \\
\left[p\right]_{\pm}&=\left[\phi\right]_{\pm}\left(\langle \mathcal A \rangle+\frac{1}{2}\langle \phi \rangle\right)+\langle \phi \rangle\left[\mathcal A\right]_{\pm}.
\end{split}
\end{equation}
Hence, using Eqs.\eqref{intschw} and \eqref{extschw} at some radius $r_H\leq R$, one finds for $\left[\phi\right]_{\pm}$ and $\left[\mathcal A\right]_{\pm}$ the following results
\begin{equation}
    \left[\phi\right]_{\pm}=\frac{2}{r_H}\left(\sqrt{1-\frac{M}{4 \pi r_H}}-\sqrt{1-\frac{M\tilde{r}_H^2}{4 \pi R^3}}\right),
\end{equation}
\begin{equation}
\begin{split}
    \left[\mathcal A\right]_{\pm}&=\frac{M}{4 \pi r_H^2\sqrt{1-\frac{M}{4 \pi r_H}}}\\
    &+\frac{M\tilde{r}_H}{4 \pi R^3\left(\sqrt{1-\frac{2M\tilde{r}_H^2}{R^3}}-3\sqrt{1-\frac{2M}{\tilde{r}_H}}\right)}.
    \end{split}
\end{equation}
From the equations above, it is clear that if the matching is performed at a radius $r_H=R$, one obtains $\left[\phi\right]_\pm=\left[\mathcal A\right]_\pm=0$, which in turn implies by Eqs.\eqref{jcLRS-Iischw} that $\bar \mu=\bar p=\bar\Pi=0$, i.e., the matching is smooth. However,  we note that this is only a particular case of a much broader class of possible outcomes. Indeed, in general, $r_H\neq R$ and the matching features a thin shell.

One can also verify that the last equation of Eqs.\eqref{jcLRS-Iischw} is always identically satisfied if  $\left[p\right]_\pm=-p^-\left(r_H\right)$. In particular, when the matching is performed at $r_H=R$, one has $p\left(r_H\right)=0$ and thus $\left[p\right]_\pm=0$. Thus, one concludes that the case for a smooth matching corresponds to a situation where the pressure is continuous across the hypersurface $H$.

Finally, consider the junction condition for $\left[\mathcal E\right]_{\pm}$, i.e., the fourth of Eq.\eqref{jcLRS-Iischw}. Since 
\begin{equation}
\begin{split}
\left[\mu\right]_{\pm}&=-\mu_0\\
\left[\mathcal E\right]_{\pm}&=\frac{M}{4\pi R^3},\\
\langle \phi \rangle[\phi]_{\pm}&=\frac{M}{2 \pi r_H^3}.
\end{split}
\end{equation}
this equation at a given radius $r_H$ reduces to
\begin{equation}\label{schwstarjcE}
\frac{M}{4\pi R^3}-\frac{M}{4 \pi r_H^3}-\frac{\mu_0}{3}=0.
\end{equation}
which in general is {\it not} satisfied, coherently with the presence of a shell. In the particular case $r_H=R$, for which the matching is smooth, this equation is identically satisfied\footnote{While this result might not be considered problematic in GR, this feature indicates that there might be some subtleties in junction conditions that have not been considered before, particularly in alternative theories of gravity connected with the junction of the Weyl tensor. For example, the solutions found in Ref.\cite{rosafluid}, where the matching was performed at a radius $r_H\neq R$, can only exist in theories of gravity for which the field equations do not depend explicitly on the Weyl tensor, one such example being GR. In more complicated theories of gravity on which the Weyl tensor plays a role in the field equations, and consequently featuring junction conditions constraining the continuity of $\mathcal E$, these solutions would not satisfy this junction condition and would not be acceptable. Another example can be found in \cite{Luz:2019frs}.}.

\subsection{The Oppenheimer-Snyder collapse}

Finally, let us now turn to the analysis of a non-static case: the Oppenheimer-Snyder collapse \cite{Oppenheimer:1939ue}. We work on this problem by making two different choices of the exterior metric to illustrate an inductive application of junction conditions and highlight the role of the observer.

\subsubsection{Dynamical exterior metric}
In this case, that matches the original approach to the problem in \cite{Oppenheimer:1939ue}, the interior solution is represented by a collapsing closed Friedmann-Lemaître-Robertson-Walker (FLRW) spacetime with line element
\begin{equation}
ds_-^2=-d\eta^2+a\left(\eta\right)^2\left(d\chi^2+\sin^2\chi d\tilde{\Omega}^2\right).\label{intOS1}
\end{equation}
For the exterior region, we consider a metric represented by a Lema\^{\i}tre-Tolman (LT) line element 
\begin{equation}
ds_+^2=-dt^2+e^{\bar\omega(t,r)}d r^2+ e^{\omega(t,r)}d\Omega^2.\label{extOS1}
\end{equation}
We also assume that the interior spacetime contains a pressureless fluid, i.e., $p^-=\Pi^-=0$ with no fluxes, i.e., $Q^-=0$ (we are then considering a comoving observer) and the exterior spacetime is empty of matter.

We consider as boundary a surface of comoving radius $\chi=\chi_H$ whose normal is given by 
\begin{equation}
n_a d\tilde{x}^a=a d\chi,
\end{equation}
which is spacelike. Choosing $e_a$ parallel to $n_a$ we have
\begin{equation}
e_a^-d\tilde{x}^a=n_a d\tilde{x}^a=a d\chi,
\end{equation}
and with the usual assumptions on $u_a^-$, we can set
\begin{equation}
u_a^- d\tilde{x}^a=-d\eta,
\end{equation}
and the congruence of the interior spacetime represents a comoving LRS observer.  Equation \eqref{n-n} implies that 
\begin{equation}
e_a^+dx^a=\frac{\partial\chi}{\partial r}e_1^-=e^{\frac{1}{2}\omega(t,r)}d r.
\end{equation}
For $u_a^+$ we have
\begin{equation}
u_a^+ dx^a=\frac{\partial\eta}{\partial t}u_0^-=-d t,
\end{equation}
so that the observer in the exterior spacetime is a comoving LRS observer. 

The tensor $q_{ab}$ in $H$ as seen from the two spacetime regions can
be described by the following line elements as
\begin{equation}\label{hOSint1}
ds_{H,-}^2=-d\eta^2+a\left(\eta\right)^2\sin^2\chi_H d\tilde{\Omega}^2,
\end{equation}
\begin{equation}\label{hOSext1}
ds_{H,+}^2=-dt^2+e^{\bar\omega(t,r_H)}d\Omega^2=-d\eta^2+e^{\omega(\eta,r_H)}d{\Omega}^2,
\end{equation}
where $r_H=r(\chi_H)$. The type I junction condition, Eq. \eqref{JC_1}, implies $t=\eta$ and
\begin{equation}\label{JC1OS1}
a\left(\eta\right)\sin\chi_H=e^{\frac{1}{2}\omega(\eta,r_H)}
\end{equation}
if $d\tilde{\Omega}_H^2=d\Omega_H^2$ i.e. if we assume the angular coordinates coincide on $H$.

As the congruences $u_a^+$, $e_a^+$, $u_a^-$, and $e_a^-$ are rotation-free and have zero expansion and shear, the remaining junction conditions are given by 
\begin{equation}\label{junctionOS}
\begin{split}
\bar\mu&=-\left[\phi\right]_{\pm}, \\
\bar p&=\frac{1}{3}\left[\phi\right]_{\pm}+\frac{2}{3}\left[\mathcal A\right]_{\pm},\\
\bar \Pi&=-\frac{1}{3}\left[\phi\right]_{\pm}-\frac{2}{3}\left[\mathcal A \right]_{\pm},\\
\left[\Sigma\right]_{\pm}&=\frac{2}{3}\left[\theta\right]_{\pm}, \\
\left[\mathcal E\right]_{\pm}&=\frac{1}{3}\left[\mu\right]_{\pm}+\frac{1}{2}\langle \phi \rangle\left[\phi\right]_{\pm}, \\
\left[p\right]_{\pm}+\left[\Pi\right]_{\pm}&=\left[\phi\right]_{\pm}\left(\langle \mathcal A \rangle+\frac{1}{2}\langle \phi \rangle\right)+\langle \phi \rangle\left[\mathcal A\right]_{\pm}.
\end{split}
\end{equation}
Let us now assume that the matching is smooth. This implies that the matter quantities at the boundary, i.e., $\bar \mu$, $\bar p$, and $\bar \Pi$, vanish. From the first of Eq. \eqref{junctionOS}, this forces the $\left[\phi\right]_\pm=0$. Then, from the second and third of the same equation, this implies further that $\left[\mathcal A\right]_\pm=0$. 

Using Eq. \eqref{JC1OS1}, one can show that the condition on the acceleration is satisfied. Next, we consider the condition on the shear and the expansion
\begin{equation}
    \left[\Sigma\right]_{\pm}=\frac{2}{3} \left[\theta\right]_{\pm} \quad \Rightarrow \quad \left. \frac{1}{2}\omega_{,\eta}\right|_{H} = \frac{a_{,\eta}}{a} ,
\end{equation}
which is satisfied if by Eq. \eqref{JC1OS1}. The fifth and the sixth of Eqs.~\eqref{junctionOS}, which concerns the electric part of the Weyl tensor, is also satisfied once the \eqref{JC1OS1} and the gravitational field equations are employed. 

The condition of the potential $\phi$ leads instead to
\begin{equation}\label{cond_phi_OS1}
   \left. \left[\phi\right]_{\pm}=0 \quad \Rightarrow \quad e^{-\frac{1}{2} \bar\omega} \omega_{,r}\right|_{H}=\frac{2 \cot \chi_H}{a}.
\end{equation}
Using Eq. \eqref{JC1OS1}, Eq. \eqref{cond_phi_OS1} leads to
\begin{equation}
4  \cos\chi (r) e^{\bar\omega} = e^{\omega}\left(\omega_{,r}\right)^2.
\end{equation}
which gives the relation between $\bar\omega$ and $\omega$ which is required for the junction.
The above equation admits as a particular solution the Oppenheimer Snyder spacetime  \cite{Oppenheimer:1939ue}
\begin{equation}
\begin{split}
e^{\bar\omega(t,r)}&=\frac{r}{\left(r^{3/2}+\frac{3  \sqrt{r_s}}{2}\eta \right)^{2/3}},\\
e^{\omega(t,r)}&=\left(r^{3/2}+\frac{3  \sqrt{r_s}}{2}\eta\right)^{4/3},
\end{split}
\end{equation}
if one chooses $\cos\chi (r)=1$. Here $r_s=\frac{1}{3}\mu_0 \sin^3\chi_H$ and $\mu_0$ is the matter energy density value in the interior spacetime at a reference instant.

\subsubsection{Static exterior metric}
In the previous section, we chose a dynamic exterior metric. This was suggested by the original work by Oppenheimer and Snyder. However, in literature, it is often found that the exterior metric of the Oppenheimer-Snyder collapse is the Schwarzschild metric. We show that the results derived above imply that the junction, in this case, is impossible for a comoving observer. However, as we have seen, we can consider a non-comoving observer {\em within} the Schwarzschild spacetime by using the tilted type II junction conditions. 

Let us start as usual by defining the interior and exterior spacetimes metric. We have 
\begin{equation}
ds_-^2=-d\eta^2+a\left(\eta\right)^2\left(d\chi^2+\sin^2\chi d\tilde{\Omega}^2\right),\label{intOS2}
\end{equation}
and 
\begin{equation}
\begin{split}
ds^2_+&=-F(R)dT^2+\frac{dR^2}{F(R)}+R^2d\Omega^2,\\
&F(R)=1-\frac{M}{4 \pi R}.
\end{split}
\label{extOS2}
\end{equation}
Again, we assume the interior spacetime contains a pressureless fluid ($p^-=\Pi^-=0$). For the exterior spacetime, we assume no matter is present. As before, we consider as boundary a surface of comoving radius $\chi=\chi_H$ whose normal is given by 
\begin{equation}
n_a d\tilde{x}^a=a d\chi,
\end{equation}
which is spacelike, and we choose $e_a$ parallel to $n_a$ 
\begin{equation}
e_a^-d\tilde{x}^a= n_a d\tilde{x}^a= a d\chi,
\end{equation} 
so that
\begin{equation}
u_a^- d\tilde{x}^a=-d\eta.
\end{equation} 
This choice selects a comoving LRS observer in interior spacetime. This time Eq. \eqref{n-n} implies that 
\begin{equation}
\begin{split}
e_0^+ \frac{dT}{d\eta} + e_1^+ \frac{dR}{d\eta}  
&=e_0^-=0,\\
e_0^+  \frac{dT}{d\chi}  + e_1^+  \frac{dR}{d\chi} &=e_1^-= a ,
\end{split}
\end{equation} 
so that
\begin{equation}
\begin{split}
e_0^+ &=  \frac{a R_{,\eta}}{R_{,\eta}T_{,\chi}-T_{,\eta}R_{,\chi}},\\
e_1^+ &=  -\frac{a T_{,\eta}}{R_{,\eta}T_{,\chi}-T_{,\eta}R_{,\chi}}.
\end{split}
\end{equation} 
For later convenience, we can set
\begin{equation}\label{SubRel}
R_{,\eta}T_{,\chi}-T_{,\eta}R_{,\chi}= -a ,
\end{equation} 
so that we can write
\begin{equation}
e_a^+d {x}^a_+= -R_{,\eta}dT + T_{,\eta} dR,
\end{equation} 
and 
\begin{equation}
u_a^+ d{x}^a_+= -F T_{,\eta}dT + \frac{R_{,\eta}}{F} dR,
\end{equation}
which does {\it not} represent an LRS comoving observer. The tensor $q_{ab}$ in $H$, as seen from the two spacetime regions, can be written in terms of the following  line elements as 
\begin{equation}\label{hOSint2}
ds_{H,-}^2=-d\eta^2+a\left(\eta\right)^2\sin^2\chi_H d\tilde{\Omega}^2,
\end{equation}
\begin{equation}\label{hOSext2}
\begin{split}
ds_{H,+}^2&=-F^2T_{,\eta}^2 dT^2+2 R_{,\eta}T_{,\eta}dRdT\\
&~~~-\frac{R_{,\eta}^2}{F^2} dR^2+R^2d\Omega^2=\\
&=-\left(F T_{,\eta}^2-\frac{R_{,\eta}^2}{F}\right)^2d\eta^2+\\
&-2\left(F T_{,\eta}T_{,\chi}-\frac{R_{,\eta}R_{,\chi}}{F}\right) d\eta d\chi\\
&+ \left(F T_{,\eta}T_{,\chi}-\frac{R_{,\eta}R_{,\chi}}{F}\right)^2 d\chi^2\\
&+R^2d\Omega^2,
\end{split}
\end{equation}
using Eq. \eqref{SubRel} for the coefficient in front of $d\eta d\chi$ and $d\chi^2$ we obtain
\begin{equation}
\begin{split}
&F T_{,\eta}T_{,\chi}-\frac{R_{,\eta}R_{,\chi}}{F}=\\
&\frac{1}{R_{,\eta}}\left(\left[F T_{,\eta}^2-\frac{R_{,\eta}^2}{F}\right]R_{,\chi}-a F T_{,\eta}\right).
\end{split}
\end{equation}
This quantity must be zero, as the induced metric must be orthogonal to $n_a$.  In this way, on $H$, the junction condition in Eq. \eqref{JC_1} implies 
\begin{equation}\label{OS2q}
\begin{split}
&\frac{R_{,\eta}^2}{F}- F T_{,\eta}^2=-1,\\
&R = a \sin\chi_H,\\
&R_{,\chi} =a F T_{,\eta},
\end{split}
\end{equation}
so that the spacelike vector $e_a^+$ is normalized (at least on $H$), and we must have 
\begin{equation}
F T_{,\eta}=\cos \chi_H= \kappa_0.
\end{equation}
Finally, using the above relations, Eq. \eqref{SubRel} becomes
\begin{equation}
T_{,\chi} = a\frac{R_{,\eta}}{F}.
\end{equation}
Considering the properties of the congruences associated with $u_a^+$, $e_a^+$, $u_a^-$, and $e_a^-$ are vorticity and twist-free, the type II junction conditions to be satisfied are 
\begin{equation}\label{junctionOS2_no}
\begin{split}
\bar\mu&=-\left[\phi\right]_{\pm}, \\
\bar p&=\frac{1}{3}\left[\phi\right]_{\pm}+\frac{2}{3}\left[\mathcal A\right]_{\pm}\\
\bar \Pi&=-\frac{1}{3}\left[\phi\right]_{\pm}-\frac{2}{3}\left[\mathcal A \right]_{\pm},\\
\left[\Sigma\right]_{\pm}&=\frac{2}{3}\left[\theta\right]_{\pm}, \\
\left[\mathcal E\right]_{\pm}&=\frac{1}{3}\left[\mu\right]_{\pm}+\frac{1}{2}\langle \phi \rangle\left[\phi\right]_{\pm}, \\
0&=\left[\phi\right]_{\pm}\left(\langle \mathcal A \rangle+\frac{1}{2}\langle \phi \rangle\right)+\langle \phi \rangle\left[\mathcal A\right]_{\pm}.
\end{split}
\end{equation}
However, as the congruence we have obtained with the type I junction conditions does not correspond to a comoving LRS observer,  we have to use tilted type II junction conditions. Indeed, even if we would not realize that the tilted conditions were necessary, one can check that the condition on the shear and expansion in Eqs.~\eqref{junctionOS2_no} can never be satisfied. Consequently, the worldline of a comoving observer of the interior spacetime cannot be smoothly matched to the static or free-falling observer in the exterior one. 

We check then if the junction is possible for a tilted observer in the exterior spacetime. In other words, we look for a parameter $\beta$ such that the following conditions are satisfied:
\begin{equation}\label{junctionOS2}
\begin{split}
\bar\mu&=-\left\{\phi\right\}_{\pm}, \\
\bar p&=\frac{1}{3}\left\{\phi\right\}_{\pm}+\frac{2}{3}\left\{\mathcal A\right\}_{\pm},\\
\bar \Pi&=-\frac{1}{3}\left\{\phi\right\}_{\pm}-\frac{2}{3}\left\{\mathcal A \right\}_{\pm},\\
\left\{\Sigma\right\}_{\pm}&=\frac{2}{3}\left\{\theta\right\}_{\pm}, \\
\left\{\mathcal E\right\}_{\pm}&=\frac{1}{3}\left\{\mu\right\}_{\pm}+\frac{1}{2}\langle \phi \rangle\left\{\phi\right\}_{\pm}, \\
\left\{p\right\}_{\pm}+\left\{\Pi\right\}_{\pm}&=\left\{\phi\right\}_{\pm}\left(\langle \mathcal A \rangle+\frac{1}{2}\langle \phi \rangle\right)+\langle \phi \rangle\left\{\mathcal A\right\}_{\pm}.
\end{split}
\end{equation}

As before, a smooth matching implies immediately $\left\{\phi\right\}_{\pm}=0$ and $\left\{\mathcal A\right\}_{\pm}=0$. From the first condition, we have
 \begin{equation}\label{phi_OS2}
    \left\{\phi\right\}_{\pm}=0 \quad \Rightarrow \quad \frac{\sqrt{F}}{R} \cosh \beta = \frac{ \cot \chi_H}{a},
\end{equation}
that using the second and the third of Eqs.\eqref{OS2q} leads to
\begin{equation}\label{Coshb}
\cosh \beta = \frac{\kappa_0}{\sqrt{F}}=\sqrt{F}T_{,\eta }.
\end{equation}
On the other hand, the conditions on the expansion and shear lead to
\begin{equation}
\begin{split}
\left\{\Sigma\right\}_{\pm}=\frac{2}{3}  \left\{\theta\right\}_{\pm} \quad \Rightarrow  \quad \frac{\sqrt{F}}{R}\sinh \beta = \frac{a_{,\eta}}{a},
\end{split}
\end{equation}
 which, combined with Eqs.\eqref{OS2q} , give
 \begin{equation}\label{Sinhb}
     \sinh \beta = \frac{R_{,\eta }}{\sqrt{F}}.
 \end{equation}
Notice that, with the results Eq.~\eqref{Sinhb} and Eq.~\eqref{Coshb}, the first of Eqs.~\eqref{OS2q}  corresponds to the well known identity
 \begin{equation}
  \cosh^2 \beta - \sinh^2 \beta=1.
 \end{equation}

Moreover, we have, for the acceleration,
\begin{equation}
\begin{split}
    \left\{\mathcal{A}\right\}_{\pm}=0 & \quad \Rightarrow \\  \frac{1}{\sqrt{F}}&\left(F_{,R}+2 {\beta}_{,\eta}\right) \cosh \beta+ 2\sqrt{F}\beta_{,\chi}\sinh \beta=0,
\end{split}
\end{equation}
which can be used to determine the value of $\beta_{,\chi}$ once ${\beta}_{,\eta}$ is known.
From Eq.~\eqref{Coshb} and Eq.~\eqref{Sinhb} we have 
\begin{equation}
{\beta}_{,\eta}=\kappa_0 \frac{F_{, R}}{F},
\end{equation}
and therefore
\begin{equation}
\beta_{,\chi}=\frac{\kappa _0 \left(F-2 \kappa _0\right) F_R}{2 F^2 \sqrt{\kappa _0^2-F}}.
\end{equation}

We now need to check only the last two equations of Eqs.\eqref{junctionOS2}.
The fifth of Eqs.\eqref{junctionOS2} implies that 
\begin{equation}
R_{,\eta\eta}=\frac{1}{2}F_R=\frac{F-1}{2 R}.
\end{equation}
which can be verified considering the derivative of the normalization constraint given by the first of Eqs.~\eqref{OS2q}. By substituting the second of Eqs.~\eqref{OS2q}  in the above equations and the form of the function $F$ one has
\begin{equation}
a_{,\eta\eta}=-\frac{M}{8\pi a^2 (1-\kappa_0^2)^{3/2} },
\end{equation}
which compared with the Raychaudhuri equation for the internal Friedmann metric 
\begin{equation}
a_{,\eta\eta}=-\frac{\mu_0}{6 a^3},
\end{equation}
gives
\begin{equation}
M=\frac{4}{3}\pi\mu_0 (1-\kappa_0^2)^{3/2}=\frac{4}{3}\pi\mu R^{3}.
\end{equation}
This last expression shows the mass of the matter of the interior spacetime as measured by the observer in the exterior spacetime.

Finally, the sixth of Eqs.\eqref{junctionOS2} is satisfied when all the previous results are employed.

We have, therefore, proven that a class of tilted observers exists for which the junction is possible. These observers are characterized by the congruences
\begin{equation}
\begin{split}\label{boost_u_e}
\breve{u}^a&= \sqrt{F}T_{,\eta } {u}^a +   \frac{R_{,\eta }}{\sqrt{F}} {e}^a, \\
\breve{e}^a&= \sqrt{F}T_{,\eta } {e}^a +  \frac{R_{,\eta }}{\sqrt{F}} {u}^a.
\end{split}
\end{equation}

\section{Conclusions}\label{sec:concl}

In this paper, we used the 1+1+2 covariant formalism to give a complete map of the junction conditions on spacelike, timelike, and null boundaries for LRS spacetimes for comoving and tilted observers. By performing the foliation at the base of the covariant formalism in such a way that $u_a$, $e_a$, or their null linear combination at the boundary hypersurface coincide with the normal, one is able to write the junction conditions in terms of the 1+1+2 potentials associated with the congruences tangent to $u_a$ and $e_a$. 

The covariant formalism offers a new perspective on junction conditions. They can be seen as the process of connecting the word lines of observers in the two spacetimes much in the same way as the interface conditions in electromagnetism.  Because of the structure of the 1+1+2 formalism, a special role is played by the motion of these observers with respect to the comoving LRS observers, i.e., the class of observers that are comoving with the matter sources and choose a space direction aligned in every point to the local axe of symmetry of the spacetime. From this point of view, junctions are only valid for two specific classes of observers in two defined geometries, and, therefore, phrases like performing the junction between the FLRW and Schwarzschild spacetimes are incomplete if no information on the observers we consider is included. Naturally, in the classical Israel treatment, the choice of the observer is performed automatically as the coordinates are chosen. Still, we feel that this way of looking at junction conditions is not as straightforward as it might lead to the idea that two spacetimes can (not) be matched {\it in absoluto}.

The 1+1+2 junction conditions have been divided into two main groups. The first ones, which we have called Type I junction conditions, do not depend on the characteristics of the two spacetimes to join and ensure the consistency of the distribution formalism associated with the junction conditions, as well as the existence of a connection between the two congruences. When covariance is broken, these conditions determine the relation between the coordinates that characterize the two spacetimes.  A second group of junction conditions, which we have called type II, are instead closely connected with the features of the spacetimes and constraint, explicitly, the features of the observers' congruences and the boundary. When covariance is broken, these add additional conditions on the parameters of the spacetime to join, if any. 

Indeed, some of these prescriptions can be quite general. For example, a consistent junction of LRS-I and LRS-II spacetimes is only possible if the scalar vorticity $\Omega$ and the twist scalar $\xi$ are continuous across the boundary for comoving LRS observers. In other words, we can only perform homologous junctions of these spacetimes. The situation changes somewhat when one considers non-comoving observers, for which it appears clear that a junction between tilted  LRS class I and LRS class III (but not class II) is possible. 

Another general result concerns the jump of the magnetic part of the Weyl tensor $\mathcal H$. We have found that in a consistent junction, this quantity must always be zero in LRS-II spacetimes but not in the other LRS classes. Therefore, vorticity and twist can play a role in the interaction of $\mathcal H$ with a boundary. As  $\mathcal H$ is associated with gravitational radiation, our result implies that one could construct spacetimes in which gravitational waves are present only in one of the components of spacetime.  In other words, gravitational radiation can be ``absorbed'' and/or ``stored'' in the vortical degrees of freedom. This possibility certainly deserves a more detailed study, which will be left for future works. 

We conclude by considering the relation between our junction conditions and the classical Israel conditions. In particular, we can explicitly prove that our formalism includes such conditions. This can be seen simply using the formulas in Appendix \ref{ExtCurv112}. For example, in the case of observers at rest in two static spherically symmetric LRS-II spacetime with spacelike boundary, the results in Table \ref{tab:spacelike}  show that the conditions for a smooth junction are essentially $\left[\phi\right]_{\pm}=0$ and $\left[\mathcal A\right]_{\pm}$ (see also Eqs. \eqref{jcLRS-Iischw}) which are consistent with setting, with the same assumptions,
\begin{equation}
\left[\mathcal{K}_{ab}^{(1)}\right]_{\pm}=0
\end{equation}
One difference, however, concerns the practical calculations. As the jumps have to be calculated in the working coordinate system, performing the explicit calculations requires less effort when dealing with scalars rather than higher-order tensors. 
Even in the case of null boundaries, for which it is now known that the Israel conditions are not sufficient to characterize completely the junction conditions \cite{clarkedray}, the additional invariants required in this case can be written in terms of 1+1+2 potentials (see Appendix \ref{Null_K} for details). This should not be surprising as these quantities characterize the complete geometry of a given spacetime. 

However, the most important difference between the covariant junction conditions and the Israel formulation is in the relevance of the conditions on the Weyl tensors for junction conditions. As Israel's conditions do not explicitly contain the Weyl tensor, these conditions might be overlooked, but they can become relevant in other settings (particular geometries or modified theories of gravity). Indeed, the ease of generalization is another advantage of using covariant approaches in formulating junction conditions. It is known that the treatment of complicated spacetimes, like Bianchi IX and extensions of General Relativity, can be significantly simplified by employing this formalism. We expect the same will happen with the extensions of junction conditions to these contexts. We will explore these possibilities in forthcoming works. 

\begin{acknowledgments}
JLR acknowledges the European Regional Development Fund and the programme Mobilitas Pluss for financial support through Project No.~MOBJD647, project No.~2021/43/P/ST2/02141 co-funded by the Polish National Science Centre and the European Union Framework Programme for Research and Innovation
Horizon 2020 under the Marie Sklodowska-Curie grant agreement No. 94533, Fundação para a Ciência e Tecnologia through project number PTDC/FIS-AST/7002/2020, and Ministerio de Ciencia, Innovación y Universidades (Spain), through grant No. PID2022-138607NB-I00.
The work of SC has been carried out in the framework of activities of the INFN Research Project QGSKY.
\end{acknowledgments}

\appendix

\section{LRS spacetimes in coordinates}\label{1+1+2 coord}
We now provide some useful expressions to calculate the 1+1+2 scalars in terms of coordinates. We have excluded the electric part of the Weyl tensor here due to its length. Its expression, however, can be derived from the definitions.

The most generic metric representing an LRS-I spacetime can be written in a coordinate system $x^a=\left(t,x,y,z\right)$ in the form \cite{Ellis:1966ta,Stewart:1967tz}
\begin{equation}
\begin{split}
    ds^2 &= -A^{-2}(t,x)[dt + E (y,k)dz]^2 \\
    &+ dx^2+C^2(x)\left[ dy^2+D^2(y,k)dz^2\right]\ , \label{LRSds}
\end{split}
\end{equation}
where $E(y,k) = (2\cos y\ , -y^2\ , -2\cosh y)$ and  $D(y,k) = (\sin y\ , y\ , \sinh y)$  for $k = (1\ , 0\ , -1)$, labeling the closed, flat, or open geometry of the 2-spaces, and $C$ is a generic function of $x$. The non-zero 1+1+2 potentials in the LRS-I case are then
\begin{align}
    \phi &= 2 \frac{C_{,x}}{C}\ ,\\
    \mathcal{A} &= -\frac{A_{,x}}{A}\ , \\
    \Omega &=-\frac{ {C}^2 {D}^2
   }{A}E_{,y},\\
   \mathcal{H} &=\frac{ E_{,x}}{A C^2 D} \left(\frac{ A_{,x}}{A}-\frac{ C_{,x}}{C} \right),
\end{align}
where commas denote the partial derivative operation.

In the LRS-II case, the most generic metric in the same coordinate system can be written as \cite{betschart}
\begin{equation}
\begin{split}
    ds^2 &= -A^{-2}(t,x)\,dt^2 + B^2(t,x)\,dx^2 \\
    &+C^2(t,x)\,[\,dy^2+D^2(y,k)\,dz^2\,]\ . \label{LRSds}
\end{split}
\end{equation}
In this case, the non-zero 1+1+2 potentials are 
\begin{align}
    \phi &= 2 \frac{ C_{,x}}{B C}\ ,\label{cphi}\\
    \mathcal{A} &= -\frac{ A_{,x}}{A B}\ , \label{cacc}\\
    \theta &= A\left(\frac{ B_{,t}}{B}+2\frac{ C_{,t}}{C}\right)\ ,\label{ctheta}\\
    \Sigma &= \frac{2}{3} A\left(\frac{ B_{,t}}{B}-\frac{ C_{,t}}{C}\right)\
    .\label{csigma}
\end{align} 

Finally, in the LRS-III case, one has a generic metric written in the form \cite{Ellis:1966ta,Stewart:1967tz}
\begin{equation}
\begin{split}
    ds^2 &= -A^{-2}(t,x)dt^2 + B^2(t)[dx-E^2(y,k) dz]^2   \\
    &+C^2(t)\left[ dy^2+ D^2(y,k)dz\right]\ , \label{LRSds}
\end{split}
\end{equation}
 and the non-zero 1+1+2 potentials are
\begin{align}
    \theta &= A\left(\frac{ B_{,t}}{B}+ 2\frac{ C_{,t}}{ C}\right)\ ,\\
    \Sigma &= \frac{2A}{3}\left(\frac{ B_{,t}}{B}- \frac{ C_{,t}}{C}\right)\ , \\
    \mathcal{A}&=-\frac{A_{,x}}{A B},\\
      \mathcal{H} &=\frac{A F_{,y}}{B C D} \left(\frac{ C_{,t}}{C}-\frac{ B_{,t}}{B} \right).
\end{align}

\section{A parametric formalism for the normal and the induced metric} \label{Normal}
Let us now construct two general vectors $v^a$ and $w^a$ defined in terms of one parameter $\epsilon$ that allows one to recover the vectors $u_a$ and $e_a$ for the particular cases $\epsilon=\pm 1$ and the basis of vectors $l_a$ and $\bar l_a$ for the particular case $\epsilon=0$. To do so, we define $v^a$ and $w^a$ as
\begin{equation}\label{defvw}
v^a=\alpha_+u_a+\alpha_0e_a,\quad w^a=\alpha_0u_a+\alpha_-e_a,
\end{equation}
\begin{equation}
\alpha_\pm=\epsilon^2\pm\alpha_0,\quad \alpha_0=\frac{\sqrt{2}}{2}\left(1-\epsilon^2\right).
\end{equation}
The vectors $v^a$ and $w^a$ defined above satisfy the following properties for the inner product
\begin{equation}\label{definnervw}
v_av^a=-\epsilon^2,\quad w_aw^a=\epsilon^2,\quad v_aw^a=\epsilon^2-1.
\end{equation}
Let us then define a set of useful matrices to work with. Define the vector matrix $V^a$, the product matrix $A$, and the projection matrix $P$ as
\begin{equation}
V^a=
\begin{bmatrix}
v^a \\ w^a
\end{bmatrix},
\end{equation}
\begin{equation}
A=
\begin{bmatrix}
-\epsilon^2 & \epsilon^2-1 \\
\epsilon^2-1 & \epsilon^2
\end{bmatrix},
\end{equation}
\begin{equation}
P=
\begin{bmatrix}
x_+ & 0 \\
0 & -x_-
\end{bmatrix},
\end{equation}
where the constants $x_\pm$ are given by
\begin{equation}
x_\pm=\frac{1}{2}\epsilon\left(1\pm\epsilon\right).
\end{equation}
Under these definitions, the 4-metric $g_{ab}$ and the induced metric $q_{ab}$ on the hypersurface take the forms
\begin{equation}\label{defgabgen}
g_{ab}=N_{ab}+V^T_a A V_b
\end{equation}
\begin{equation}\label{defhabgen}
q_{ab}=N_{ab}+V^T_a P A V_b\equiv N_{ab}+V^T_a B V_b,
\end{equation}
where we have defined a matrix $B=PA$ as the product matrix for the 3-dimensional hypersurface.

\subsubsection{The case $\epsilon = \pm1$: timelike and spacelike hypersurfaces}

Let us assume $\epsilon = \pm 1$. In this case, one has $\alpha_0=0$, $\alpha_\pm=1$. From Eq. \eqref{defvw}, one verifies that the vectors $v^a$ and $w^a$ reduce to  $u_a^H$ and $e_a^H$, respectively, and that the inner products in Eq.\eqref{definnervw} reduce to 
\begin{equation}
  g_{ab} u^a  u^b =-1, \quad  g_{ab}e^a e^b=1 \quad g_{ab}u^a e^b =0.
\end{equation}
in $H$. Furthermore, the product matrix $A$ reduces to 
\begin{equation}
    A=\text{diag}\left(-1,1\right)
\end{equation}
and thus the 4-metric $g_{ab}$ reduces to the form given in Eq.\eqref{metricin112}.

For timelike hypersurfaces, i.e., for $\epsilon=1$, the constants $x_\pm$ reduce to $x_+=1$ and $x_-=0$. The projection matrix is \begin{equation}
   P=\text{diag}\left(1,0\right) 
\end{equation} and one obtains $B=-P$. This implies that the normal vector satisfies $n_a=e^H_a $  and  the induced metric $q_{ab}$ in the hypersurface is $h_{ab}$ given by Eq.\eqref{metricin112} in $H$:
\begin{equation}
    q _{ab}= N^H _{ab}-u^H_{a} u^H_{b} .
\end{equation}
On the other hand, for spacelike hypersurfaces, i.e., for $\epsilon=-1$, the constants $x_\pm$ become instead $x_+=0$ and $x_-=-1$.  The projection matrix becomes \begin{equation}
    P=\text{diag}\left(0,1\right)
\end{equation} 
and also $B=P$. As a result, the normal vector  can be chosen as $n_a=u^H_a$ and induced metric $q_{ab}$ becomes
\begin{equation}
    q _{ab}= N^H_{ab}+e^H_{a} e^H_{b} .
\end{equation}

\subsubsection{The case $\epsilon = 0$: null hypersurfaces}

Let us now set $\epsilon=0$. In this case, one obtains $\alpha_+=-\alpha_-=\alpha_0=\sqrt{2}/2$. From Eq. \eqref{defvw}, one concludes that the vectors $v^a$ and $w^a$ become the null vectors $l_a$ and $\bar l_a$ respectively, and the inner products from Eq.\eqref{definnervw} become the ones in Eq.\eqref{definnerll}:
\begin{equation}\label{defvecll}
l_a^H =\frac{\sqrt{2}}{2}\left( u_a^H +e_a^H \right), \quad \bar{l}_a^H =\frac{\sqrt{2}}{2}\left(u_a^H -e_a^H \right), 
\end{equation}
Also, in this case, the constants $x_\pm$ vanish, which implies that the projection matrix $P$, and consequently the matrix $B$, vanish identically. Thus, as anticipated, the induced metric $q_{ab}$ coincides with the 2-metric $N_{ab}$. The matrix $A$ is given by
\begin{equation}
A=\begin{bmatrix}
0 & -1 \\
-1 & 0
\end{bmatrix},
\end{equation}
from which one verifies that Eqs.\eqref{defgabnull} and \eqref{defhabnull} for $g_{ab}$ and $q_{ab}$ respectively are recovered. Furthermore, the normal vector $n_a$ becomes the null vector $\bar l_a $, and all the results are consistent.

\subsubsection{Paramentric for of the normal vector}
The discussion above proves explicitly that we can always write the normal vector $n_a$ to the hypersurface in general as 
\begin{equation}
n_a=\tau u_a+\varsigma e_a,
\end{equation}
where 
\begin{equation}
\tau=\alpha_0-x_-, \qquad \varsigma=x_+-\alpha_0,
\end{equation}
which is precisely Eq. \eqref{defnormgen}. Notice also that in this case
\begin{equation}
\varepsilon=\varsigma^2-\tau^2.
\end{equation}

\section{Derivatives of the Dirac-$\delta$ distribution}\label{sec:APPdeltaderiv}

When dealing with derivatives of the 1+1+2 scalars that feature terms proportional to the $\delta\left(\ell\right)$ distribution, e.g., the scalars $\mathcal E$, $\mathcal H$, the Gaussian curvature $K$, and the matter fields $\mu$, $p$, $\Pi$, and $Q$, it is necessary to compute the covariant derivatives of the $\delta$ distribution function. These derivatives give rise to additional contributions non-tangent to the hypersurface $H$, usually called double gravitational layers and external fluxes and tensions \cite{senovilla1,Reina:2015gxa}. In this section, we briefly review how to compute these derivatives.

Consider a scalar quantity $X$ that can be written in the distribution formalism as
\begin{equation}
    X=X^+\Theta\left(\ell\right)+X^-\Theta\left(-\ell\right)+\bar X\delta\left(\ell\right),
\end{equation}
where $\bar X$ denotes the term proportional to the $\delta$ distribution. Taking a covariant derivative of $X$, one obtains
\begin{equation}\label{eq:AppdX}
\begin{split}
    \nabla_a X=&\left(\nabla_aX\right)^+\Theta\left(\ell\right)+\left(\nabla_a X\right)^-\Theta\left(-\ell\right)+\\
    +&\left[X\right]_\pm\delta\left(\ell\right)+\nabla_a\bar X \delta\left(\ell\right)+\bar X\nabla_a\delta\left(\ell\right).
    \end{split}
\end{equation}
To compute the last term on the right-hand side of Eq.\eqref{eq:AppdX}, one needs to consider the full definition of a distribution function. Let $Y^a$ be a tensorial test function of compact support. We define the application of a distribution function of the form $\nabla_a\left(\delta\left(\ell\right)\right)$ to the test function $Y^a$ as
\begin{equation}
    \left<\nabla_a\delta,Y^a\right>=\int_\Omega \nabla_a\delta\left(\ell\right)Y^a d\Omega.
\end{equation}
Given that the test function $Y^a$ has compact support, one can perform an integration by parts to obtain
\begin{equation}
    \left<\nabla_a\delta,Y^a\right>=-\left<\delta,\nabla_a Y^a\right>.
\end{equation}
The covariant derivative $\nabla_a Y^a$ can then be split into its tangent and orthogonal projections with respect to the hypersurface $H$ via 
\begin{equation}
\nabla_a Y^a=q_a^b \nabla_b Y^a+n_a n^b\nabla_b Y^a.
\end{equation}
The projection orthogonal to $H$ can then be manipulated as follows,
\begin{equation}
    \left<\delta,n_a n^b\nabla_b Y^a\right>=-\left<\nabla_b\left(n_a n^b \delta\right),Y^a\right>.
\end{equation}
Thus, defining a distribution $\Delta_a\equiv\nabla_b\left(n_a n^b \delta\left(\ell\right)\right)$, one can interpret the projection orthogonal to $H$ as the application of some tensor distribution function $\Delta_a\left(\ell\right)$ on a test function $Y^a$:
\begin{equation}
    \left<\Delta_a,Y^a\right>=-\int_H\epsilon n_a n^b\nabla_b Y^a dH
\end{equation}
This distribution is commonly referred to as the double gravitational layer. On the other hand, the projection tangent to $H$ can be manipulated via the use of the Gauss's theorem, from which one obtains
\begin{equation}
    \left<\delta,q_a^b \nabla_b  Y^a\right>=\left<\delta,\epsilon n_a \mathcal{K}\; Y^a\right>
\end{equation}
where $\mathcal{K}$ is the trace fo the extrinsic curvature $\mathcal{K}_{ab}$ (see Sec. \ref{Null_K} for its definition). Collecting all the results above, one verifies that the covariant derivative of the $\delta$ distribution features a term proportional to the double gravitational layer plus an orthogonal term proportional to the trace of the extrinsic curvature. Summarizing,
\begin{equation}
    \nabla_a\delta\left(\ell\right)=\Delta_a\left(\ell\right)-\epsilon n_a \left<\mathcal{K}\right>\delta\left(\ell\right).
\end{equation}
We note that the double gravitational layer distribution $\Delta_a\left(\ell\right)$ corresponds to a singular term non-tangent to the hypersurface $H$. Thus, upon calculating the junction conditions in this manuscript, in the same way that the terms proportional to $\delta\left(\ell\right)$ must match on both sides of the 1+1+2 equations, also the terms proportional to $\Delta_a\left(\ell\right)$ must match on both sides of these equations, which leads to additional constraints that simplify the calculations and the system of junction conditions.

One final remark regarding the notation used in this section and throughout the manuscript for the distribution $\Delta_a\left(\ell\right)$. In other publications, one may find a slightly different definition of this distribution, given by $\Delta_a^X\left(\ell\right)=\nabla_b\left(\bar X n_a n^b\delta\left(\ell\right)\right)$, i.e., with a factor $\bar X$ inside. If one wishes to transform the notation from $\Delta_a\left(\ell\right)$ to $\Delta_a^X\left(\ell\right)$, this can be achieved through the equality
\begin{equation}
    \nabla_a\bar X+\bar X\Delta_a\left(\ell\right)=q_a^b\nabla_b\bar X+\Delta_a^X\left(\ell\right).
\end{equation}
In this work, we opt for the former notation due to its simpler structure and convenience in the analysis carried out throughout the manuscript. Still, we note that the two notations are equivalent.

\section{Second fundamental forms} \label{Null_K}

In this appendix, we show that the junction conditions obtained in Sec. \ref{sec:junctions} are indeed in agreement with the Israel junction conditions, which are based on the jump of the extrinsic curvature and the induced metric.  

\subsection{Extrinsic curvature in terms of the 1+1+2 quantities}\label{ExtCurv112}

For every type of hypersurface, i.e., timelike, spacelike, and null, the extrinsic curvature of the hypersurface is given by the projection of the covariant derivative of the normal vector
\begin{equation}\label{defkab}
\mathcal{K}_{ab}=q^c_a q^d_b\nabla_cn_d,
\end{equation}
where in general the induced metric $q_{ab}$ is given by Eq.\eqref{defhabgen}, and the normal vector is given by Eq.\eqref{defnormgen}. Equation \eqref{defkab} can be written in terms of the parametric formalism (see Appendix \ref{Normal}) as
\begin{equation}
\mathcal{K}_{ab}=x_+\mathcal{K}_{ab}^{(1)}- x_-\mathcal{K}_{ab}^{(2)}+\alpha_0\mathcal{K}_{ab}^{(3)}
\end{equation}
where we have defined
\begin{eqnarray}\label{extrinsic1}
\mathcal{K}_{ab}^{(1)}&=&-\mathcal A u_a u_b+\frac{1}{2}\phi N_{ab}+\xi\epsilon_{ab}
\end{eqnarray}
\begin{eqnarray}\label{extrinsic2}
\mathcal{K}_{ab}^{(2)}&=&\frac{1}{3}\theta\left(e_ae_b+N_{ab}\right)+\Sigma\left(e_ae_b-\frac{1}{2}N_{ab}\right)
\end{eqnarray}
\begin{equation}\label{extrinsic3}
\mathcal{K}_{ab}^{(3)}=N_{ab}\left(\frac{1}{3}\theta-\frac{1}{2}\Sigma+\frac{1}{2}\phi\right)+\epsilon_{ab}\left(\xi+\Omega\right).
\end{equation}
We can express $\mathcal{K}_{ab}$ also in terms of $\tau$ and $\varsigma$ as
\begin{equation}
\mathcal{K}_{ab}= \varsigma\mathcal{K}_{ab}^{(1)}- \tau\mathcal{K}_{ab}^{(2)}-\alpha_0\left(\mp\mathcal{K}_{ab}^{(1)}-\mathcal{K}_{ab}^{(2)}-\mathcal{K}_{ab}^{(3)}\right)
\end{equation}
where the choice of the sign in forn to of $\mathcal{K}_{ab}^{(1)}$ depends on the null vector chosen and
\begin{equation}
\alpha_0=\frac{\sqrt{2}}{2}\left[1-\left(\varsigma^2-\tau^2\right)\right]
\end{equation}
For timelike and spacelike hypersurfaces, one has $\tau=1$ and $\varsigma=0$ or $\tau=0$ and $\varsigma=1$, which in both cases implies that $\alpha_0=0$. Then, for timelike hypersurfaces, one has 
 extrinsic curvature becomes 
 \begin{equation}
 \mathcal{K}_{ab}=\mathcal{K}_{ab}^{(1)},
 \end{equation}
 and for spacelike hypersurfaces, the extrinsic curvature is
  \begin{equation}
 \mathcal{K}_{ab}=\mathcal{K}_{ab}^{(2)}.
 \end{equation}
Finally, for null hypersurfaces, $\tau=\pm\varsigma=\frac{\sqrt{2}}{2}$, $\alpha_0=\frac{\sqrt{2}}{2}$, and the extrinsic curvature reads 
  \begin{equation}
 \mathcal{K}_{ab}=\frac{\sqrt{2}}{2}\mathcal{K}_{ab}^{(3)}.
 \end{equation}

Let us now verify that the smoothness of the matching, as defined by the $1+1+2$ potentials, implies the continuity of the extrinsic curvature. For timelike hypersurfaces, one verifies that the matching is smooth whenever
\begin{equation}
\left[\theta\right]_{\pm}=\left[\Sigma\right]_{\pm}=0,
\end{equation}
which corresponds to the continuity of the extrinsic curvature as seen from Eq.\eqref{extrinsic2} in the particular case for LRS spacetimes. On the other hand, for spacelike hypersurfaces, the matching is smooth whenever 
\begin{equation}
\left[\phi\right]_{\pm}=\left[\mathcal A\right]_{\pm}=0,
 \end{equation}
which corresponds to the continuity of the extrinsic curvature from Eq.\eqref{extrinsic1} in the particular case for LRS spacetimes. Finally, for null hypersurfaces, in LRS-II spacetimes, the matching is smooth if
\begin{equation}
\left[\phi\right]_\pm=0
\end{equation}
which implies 
\begin{equation}
\left[\Sigma\right]_\pm=\frac{2}{3}\left[\phi\right]_\pm
\end{equation}
Both these conditions correspond to the continuity of the extrinsic curvature from Eq.\eqref{extrinsic3}. However, an additional constraint is necessary for smoothness, namely:
\begin{equation}
    \left[\theta\right]_\pm=\pm\left[\mathcal A\right]_\pm.
\end{equation}
Nevertheless, since $\mathcal A$ is unconstrained in this case, the previous expression simply indicates that the value of $\left[\mathcal A\right]_\pm$ determines merely the value of  $\left[\theta\right]_\pm$.

\subsection{Other fundamental forms for null geodesics}

In the case of null geodesics, it is necessary to analyze other second fundamental forms apart from the extrinsic curvature (see, e.g., \cite{clarkedray}). In particular, the following fundamental forms have been used in the literature:
\begin{equation}
\chi_{ab}=q^c_aq^d_b\nabla_c l_d,
\end{equation}
\begin{equation}
\psi_{ab}=q^c_aq^d_b\nabla_c \bar l_d,
\end{equation}
\begin{equation}
\eta_a=q^c_a\bar l^d\nabla_c l_d=-q^c_a l^d\nabla_c \bar l_d.
\end{equation}
In terms of the $1+1+2$ potentials and for null hypersurfaces, the fundamental form $\eta_a$ vanishes identically, whereas the fundamental forms $\chi_{ab}$ and $\psi_{ab}$ take the forms
\begin{equation}\label{fundamental1}
\chi_{ab}=\frac{1}{\sqrt{2}}\left[N_{ab}\left(\frac{1}{3}\theta-\frac{1}{2}\Sigma+\frac{1}{2}\phi\right)+\epsilon_{ab}\left(\Omega+\xi\right)\right],
\end{equation}
\begin{equation}\label{fundamental2}
\psi_{ab}=\frac{1}{\sqrt{2}}\left[N_{ab}\left(\frac{1}{3}\theta-\frac{1}{2}\Sigma-\frac{1}{2}\phi\right)+\epsilon_{ab}\left(\Omega-\xi\right)\right].
\end{equation}
Fom the junction conditions $\left[\xi\right]_\pm=\left[\Omega\right]_\pm=0$, one verifies that the terms proportional to $\epsilon_{ab}$ in both Eqs.\eqref{fundamental1} and \eqref{fundamental2} above are continuous. Furthermore, as mentioned before, for null hypersurfaces the smoothness of the matching implies that $\left[\phi\right]_\pm$, which consequently implies that either $\left[\theta\right]_\pm=\left[\Sigma\right]_\pm=0$ for LRS-I and LRS-III spacetimes or $\left[\Sigma\right]_\pm=\frac{2}{3}\left[\theta\right]_\pm$ for LRS-II spacetimes. Both these conditions imply the continuity of the terms proportional to $N_{ab}$ in Eqs. \eqref{fundamental1} and \eqref{fundamental2}, thus implying that these fundamental forms are continuous in the particular case of smooth matching, as anticipated.


\end{document}